\documentclass[twocolumn]{aastex631}

\newcommand{\ovi}{O\,\textsc{vi}}

\usepackage{amsmath}
\usepackage{icomma}

\begin{document}

\title{The Origin of the Cluster of Local Interstellar Clouds}

\correspondingauthor{Catherine Zucker}
\email{catherine.zucker@cfa.harvard.edu}

\author[0000-0002-2250-730X]{Catherine Zucker}
\affiliation{Center for Astrophysics $\mid$ Harvard \& Smithsonian, 
60 Garden St., Cambridge, MA, USA 02138}
\affiliation{Space Telescope Science Institute, 3700 San Martin Drive, Baltimore, MD 21218, USA}

\author[0000-0003-3786-3486]{Seth Redfield}
\affiliation{Astronomy Department and Van Vleck Observatory, Wesleyan University, Middletown, CT 06459, USA}

\author[0009-0002-7155-6180]{Sara Starecheski}
\affiliation{Astronomy Department and Van Vleck Observatory, Wesleyan University, Middletown, CT 06459, USA}
\affiliation{Department of Physics, Sarah Lawrence College, Bronxville, NY 10708, USA}

\author[0000-0001-8235-2939]{Ralf Konietzka}
\affiliation{Center for Astrophysics $\mid$ Harvard \& Smithsonian, 
60 Garden St., Cambridge, MA, USA 02138}

\author[0000-0003-4446-3181]{Jeffrey L. Linsky}
\affiliation{JILA, University of Colorado and NIST, Boulder, CO 80309-0440, USA
}



\begin{abstract}
The interstellar medium within $\rm\approx 15 \; pc$ of the Sun consists of a complex of fifteen diffuse, partially ionized clouds. Located within the Local Bubble, these clouds, known as the Cluster of Local Interstellar Clouds (CLIC), constitute the interstellar environment impinging upon our heliosphere. While each individual cloud can be modeled with a distinct velocity vector, the complex demonstrates a coherent bulk motion suggestive of a common origin. Here we examine two theories for the origin of the CLIC:  that it formed due to an ionization front associated with nearby Str\"{o}mgren spheres and/or due to a nearby supernova explosion that occurred within the pre-evacuated cavity of the Local Bubble. Tracing back the trajectory of the clouds, we disfavor a purely Str\"{o}mgren sphere origin, given the CLIC's position interior to the surface of the most significant nearby Stromgren sphere and its motion transverse to the sphere's trajectory. Turning to a supernova origin, we model the formation of the CLIC assuming individual clouds have been swept up over time due to the expansion of a supernova remnant in its pressure-driven snowplow phase. We find that the 3D spatial-dynamical properties of the CLIC can be explained by the most recent supernova that exploded in the nearby Upper Centaurus Lupus cluster $\approx \rm 1.2 \; Myr$ ago and propagated into an ambient density of $n \approx 0.04 \;\rm cm^{-3}$. Our model predicts that the formation of the individual CLIC clouds occurred progressively over the past $1 \; \rm Myr$ and offers a natural explanation for the observed distribution, column density, temperature, and magnetic field structure of the complex. 
\end{abstract}



\section{Introduction} \label{sec:intro}

The local interstellar medium is dominated by a large cavity known as the Local Bubble, a region of ionized, tenuous plasma extending $\rm \approx 100-200 \; pc$ from the Sun in all directions \citep{frisch2011,pelgrims2020,Lallement2019,ONeill_2024}. The Local Bubble is embedded between two large-scale Galactic features, the oscillating Radcliffe Wave \citep{Konietzka2024, Alves2020} and the Split \citep{Lallement2019}. While the existence of the Local Bubble has been known for over four decades \citep{cox1987, lucke1978, sanders1977}, there is still debate on both the nature and origin of the gas within it.

For example, some studies argue that the Local Bubble is mostly filled with a hot-rarefied medium with typical temperatures of $\rm 10^6\,K$ \citep{Galeazzi2014}, while others argue that the Local Bubble is characterized by much cooler, but still warm, gas at temperatures of $10,000-20,000\; \rm K$ \citep{jenkins2020}. Most evidence suggests that supernovae originating in the nearby Scorpius-Centaurus OB (Sco-Cen) association are responsible for carving out its present day morphology \citep{maizapellaniz2001, fuchs2006, breitschwerdt2016} and giving rise to some fraction of the X-ray emission detected with ROSAT \citep{snowden2000}. \citet{linsky2021} proposed that properties of the Local Bubble could also be shaped by a Str\"{o}mgren sphere \citep{stromgren1939} stemming from the nearby B star $\epsilon$ CMa, which may explain why most of the cavity is fully ionized, but a deficiency in \ovi-bearing gas ($T \rm >300,000 \; K$) is detected near the Sun \citep{barstow2010}. Regardless of what temperature gas pervades the cavity, there is a consensus that the cavity is bounded by a dense shell of colder, neutral gas and dust \citep{pelgrims2020, Lallement2003, Welsh2010, ONeill_2024}, with \citet{zucker2022} arguing that all star formation within $\rm \approx 200 \; pc$ of the Sun is occurring on the Local Bubble's surface.

The gas within the Local Bubble is not uniform, and significant structure exists in the form of the Cluster of Local Interstellar Clouds (CLIC), a complex of warm ($T \approx 5,000-10,000 \; \rm K$), diffuse \citep[$n\approx \rm 0.1 \; cm^{-3}$; see e.g. Figure 14 from][]{Linsky_2022}, and partially ionized clouds that lie within $\rm \approx 15 \; pc$ from the Sun. The Sun is currently traveling through one \citep[the Local Interstellar Cloud (LIC);][]{linsky2019} or more \citep[a mixture of the LIC and G clouds;][]{Swaczyna_2022} of these clouds, which is known to have a substantial impact on the configuration of the heliosphere and the influx of Galactic cosmic rays on Earth \citep{Linsky_2022}. The properties of these clouds have largely been constrained via high-spectral-resolution optical and UV absorption line observations toward nearby stars. \citet{redfield2008} sample the physical properties of warm interstellar gas along 157 lines of sight toward stars within 100 pc of the Sun using {\it Hubble Space Telescope} {\it (HST)} spectroscopy. \citet{redfield2008} leveraged these measurements to construct a dynamical model of the CLIC that consists of 15 clouds, with each cloud possessing kinematics that can roughly be defined by a single velocity vector. 

Previous studies have proposed that the origin of the CLIC could be related either to nearby supernova-driven bubbles (e.g., the Local Bubble, Loop I) and/or to nearby Str\"{o}mgren spheres. For example, \citet{frisch2011} proposed that the CLIC clouds may be embedded in the rim of the ``S1" subshell --- a component of the Loop I magnetic superbubble \citep[see also][]{frisch2010}. \citet{frisch2017} discussed several lines of evidence for this interpretation, most notably that the upwind direction of the mean velocity vector of the CLIC is directed towards the center of the Loop I superbubble \citep[see also][]{slavin2009,frisch1981}. Other studies suggested an origin for the CLIC in the potential interaction between the Loop I superbubble and the Local Bubble \citep{breitschwerdt2000}. Either way, supernovae from the nearby Sco-Cen OB association have been implicated in the origin of the CLIC --- in addition to the origin of Loop I and the Local Bubble --- and have been proposed to explain the CLIC's bulk motion away from the Galactic center. More recently, \citet{Piecka_2024} bolstered the idea that the very local ISM is dynamically linked to a uniform gas flow originating in the Sco-Cen association, but argued that further follow up work is needed to understand the relative contributions from supernovae and/or stellar winds in shaping this flow. Finally, \citet{linsky2021} suggest that nearby Str\"{o}mgren spheres may have interacted with these recent supernova shockwaves, potentially also contributing to the CLIC's formation. 

New astrometric data from the Gaia space mission have enabled unprecedented constraints not only on the 3D spatial structure of the local interstellar medium \citep{Leike2020, Lallement2019, Edenhofer2024} and its constituent superbubbles \citep[e.g., the Local Bubble and Loop I;][]{pelgrims2020, ONeill_2024, panopoulou2021}, but also on the 3D structure and dynamics of nearby OB associations and other sources of ionizing radiation. Combining these new Gaia-based constraints with complementary data on the 3D positions and kinematics of the CLIC clouds derived from extant UV spectroscopy \citep{redfield2008} offers an unprecedented opportunity to revisit the origin of the CLIC in renewed detail. 

In this work, we explore two possibilities for the origin of the CLIC: that the CLIC formed either due to an ionization front associated with nearby Str\"{o}mgren spheres and/or due to a supernova explosion stemming from the nearby Sco-Cen OB association. 

In \S \ref{sec:clicprops}, we start by presenting the 3D spatial and dynamical constraints on the CLIC, which will underpin investigations into the potential origin of the complex tied to both nearby Str\"{o}mgren spheres and nearby supernovae. 

In \S \ref{sec:stromgren} we explore the possibility of a Str\"{o}mgren sphere origin, compiling Gaia and Hipparcos data on nearby ionizing sources --- including B stars and hot white dwarfs --- to construct their sizes and 3D distribution with respect to the CLIC. We trace the trajectory of the CLIC backward in time alongside the largest Str\"{o}mgren sphere in the solar vicinity to disfavor a potential origin for the CLIC solely related to the largest, nearby Str\"{o}mgren sphere. 

In \S \ref{sec:SN}, we explore the possibility of a supernova origin for the CLIC. We posit that the most recent supernova shockwave occurring between $\rm \approx 1-2 \; Myr$ ago in the Upper Centaurus Lupus (UCL) subgroup of the Sco-Cen OB association \citep{neuhauser2020} propagated into the pre-evacuated cavity of the Local Bubble and could have given rise to the CLIC. Given this hypothesis, we fit for the evolution of this supernova remnant using the backward trajectories of the CLIC components, under the assumption that the individual clouds may been swept up and formed at different periods in the shell’s evolution.

In \S \ref{sec:discussion} we explore the implications of our shell modeling from \S \ref{sec:SN}, including predictions for the ages and birth sites of the CLIC and possible explanations for their observed physical properties. We also discuss limitations of our modeling. Finally, we conclude in \S \ref{sec:conclusions}.

\begin{deluxetable*}{ccccccccccccccc}[ht!]
\tablecaption{Spatial and Dynamical Properties of the CLIC \label{tab:clic6d} }
\tablehead{\colhead{Cloud} & \colhead{$l_{cen}$} & \colhead{$b_{cen}$} & \colhead{$d$} & \colhead{$V_0$} & \colhead{$l_0$} & \colhead{$b_0$} & \colhead{$x$} & \colhead{$y$} & \colhead{$z$} & \colhead{$u$} & \colhead{$v$}  & \colhead{$w$} &  \colhead{$v_{mag}$} & \colhead{$\sigma v_{mag}$}\\
\colhead{} & \colhead{$^\circ$} & \colhead{ $^\circ$} & \colhead{pc} & \colhead{$\rm km \; s^{-1}$} & \colhead{deg} & \colhead{deg} & \colhead{pc} & \colhead{ pc} & \colhead{ pc} & \colhead{$\rm km \; s^{-1}$} & \colhead{$\rm km \; s^{-1}$} & \colhead{$\rm km \; s^{-1}$} & \colhead{$\rm km \; s^{-1}$} & \colhead{$\rm km \; s^{-1}$} \\ \colhead{(1)} & \colhead{(2)} & \colhead{(3)} & \colhead{(4)} & \colhead{(5)} & \colhead{(6)} & \colhead{(7)} & \colhead{(8)} & \colhead{(9)} & \colhead{(10)} & \colhead{(11)} & \colhead{(12)} & \colhead{(13)} & \colhead{(14)} & \colhead{(15)}}
\startdata
LIC & 170 & -10 & 1.1 & 23.84±0.90 & 187.0±3.4 & -13.5±3.3 & -1.1 & 0.2 & -0.2 & -13.0 & 12.6 & 2.2 & 18 & 1.3 \\
G & 315 & 0 & 1.3 & 29.6 ±1.1 & 184.5± 1.9 & -20.6 ±3.6 & 0.9 & -0.9 & 0.0 & -17.6 & 13.2 & -2.6 & 22 & 1.1 \\
Blue & 250 & -30 & 2.6 & 13.89± 0.89 & 205.5 ±4.3 & -21.7± 8.3 & -0.8 & -2.1 & -1.3 & -1.6 & 9.8 & 2.7 & 10 & 1.0 \\
Aql & 40 & -5 & 3.5 & 58.6 ±1.3 & 187.0 ±1.5 & -50.8 ±1.0 & 2.7 & 2.2 & -0.3 & -26.8 & 10.9 & -37.6 & 47 & 1.2 \\
Eri & 70 & -20 & 3.5 & 24.1± 1.2 & 196.7± 2.1 & -17.7 ±2.6 & 1.1 & 3.1 & -1.2 & -12.0 & 8.8 & 0.5 & 14 & 1.0 \\
Aur & 210 & 10 & 3.5 & 25.22±0.81 & 212.0± 2.4 & -16.4 ±3.6 & -3.0 & -1.7 & 0.6 & -10.5 & 2.6 & 0.7 & 10 & 1.0 \\
Hyades & 180 & -20 & 5.0 & 14.69±0.81 & 164.2± 9.4 & -42.8 ±6.1 & -4.7 & 0.0 & -1.7 & -0.4 & 18.3 & -2.2 & 18 & 1.4 \\
Mic & 40 & 15 & 5.1 & 28.45± 0.95 & 203.0 ±3.4 & -03.3± 2.3 & 3.8 & 3.2 & 1.3 & -16.1 & 4.3 & 6.2 & 17 & 1.2 \\
Oph & 45 & 25 & 5.1 & 32.25± 0.49 & 217.7± 3.1 & +00.8 ±1.8 & 3.3 & 3.3 & 2.2 & -15.5 & -4.3 & 8.3 & 18 & 0.9 \\
Gem & 300 & 40 & 6.7 & 36.3±1.1 & 207.2 ±1.6 & -01.2± 1.3 & 2.6 & -4.4 & 4.3 & -22.3 & -1.2 & 7.0 & 23 & 1.1 \\
NGP & 5 & 75 & 8.5 & 37.0 ±1.4 & 189.8 ±1.7 & -05.4 ±1.1 & 2.2 & 0.2 & 8.2 & -26.3 & 9.1 & 4.3 & 28 & 1.3 \\
Leo & 270 & 55 & 11.1 & 23.5± 1.6 & 191.3± 2.8 & -08.9 ±1.8 & -0.0 & -6.4 & 9.1 & -12.8 & 10.9 & 4.2 & 17 & 1.3 \\
Dor & 270 & -50 & 11.7 & 52.94± 0.88 & 157.3 ±1.5 & -47.93 ±0.63 & -0.0 & -7.5 & -9.0 & -22.7 & 29.1 & -31.5 & 48 & 0.9 \\
Vel & 300 & -45 & 14.9 & 45.2 ±1.8 & 195.4 ±1.1 & -19.1 ±1.0 & 5.3 & -9.1 & -10.5 & -31.2 & 4.1 & -7.0 & 32 & 1.7 \\
Cet & 290 & -40 & 15.5 & 60.0± 2.0 & 197.11± 0.56 & -08.72± 0.50 & 4.1 & -11.2 & -10.0 & -46.7 & -2.0 & -1.3 & 46 & 1.8 \\
\enddata
\tablecomments{The 3D positions and 3D space motions of the CLIC, derived from data taken from \citet{redfield2008} (see their Tables 16 and 18). (1) Name of the cloud (2--3) The central Galactic coordinates of the cloud (4) The distance of the closest star with the cloud’s absorption velocity, equivalent to an upper limit on the cloud's distance (5--7) The velocity magnitude ($V_0$) and the direction in Galactic coordinates ($l_0$, $b_0$) that best fits the set of radial velocity components detected in absorption toward each cloud. (8--10) The Heliocentric Galactic Cartesian coordinates of the cloud, derived from Columns 2-7. (11--13) The 3D space motions along $x$, $y$, and $z$ with respect to the LSR. (14--15) The magnitude of the 3D velocity vector of the cloud with respect to the LSR and its corresponding uncertainty. A machine readable version of this table is available online at the Harvard Dataverse (\href{https://doi.org/10.7910/DVN/CQJZYH}{doi:10.7910/DVN/CQJZYH}).}
\end{deluxetable*}

\section{3D Positions and Velocities of the CLIC Clouds} \label{sec:clicprops}
To construct the dynamical histories of the individual clouds in the CLIC, we leverage measurements of their projected sky positions and velocities, which have been well measured by the growing database of high-resolution UV spectra taken with {\it HST}. 

\citet{redfield2008} presented a kinematic and 2D morphological model of the CLIC based on 270 local interstellar absorption components toward more than 150 nearby stars. The model assumes homogeneous clouds with sharp boundaries that can be described kinematically with a single velocity vector. This model results in 15 distinct clouds that reside within approximately 15 pc of the Sun. An alternative model is presented by \citet{gry2014}, where instead of a multitude of simple kinematic clouds, they presented a model of a small number of interstellar clouds with more complex kinematic structures (e.g., shocks). \citet{redfield2015} presented a comparative analysis of new {\it HST} observations and the 15-cloud model is more effective at predicting the absorption toward nearby stars. 

Accordingly, we use the \citet{redfield2008} constraints on each cloud's heliocentric velocity vector to construct the 3D heliocentric Galactic Cartesian space motions with respect to the Local Standard of Rest (LSR), hereafter denoted ($u$, $v$, $w$).\footnote{For the solar motion, we adopt $(U_\odot, V_\odot, W_\odot) = (10.0, 15.4, 7.8) \; \rm km \; s^{-1}$ from \citet{Kerr_LyndenBell_1986}, as adopted in \citet{zucker2022}.} The \citet{redfield2008} 3D heliocentric velocity vector is defined by the velocity magnitude $(V_0)$ and the direction in Galactic coordinates ($l_0$, $b_0$) that best fits the collection of radial velocity measurements published in \citet{redfield2008}. These distinct local interstellar cloud vectors are an important observational constraint that must be explained by any theory for the origin of the CLIC. 

To model the 3D spatial distribution of the CLIC, we require an estimate of their distances. The distances to the individual clouds in the CLIC are not well constrained. An upper limit on the distance of the closest edge of the cloud can be obtained using the distance of the nearest star that shows evidence of the cloud in absorption. However, we do not know where along the line of sight the absorption occurs. Together with a large survey of nearby stars, one can reconstruct the 3D morphology of the CLIC. Initial attempts at this have been made (e.g., see Figure \ 11 in \citealt{frisch2011} and \citealt{vannier2019}). In this work, we model the 3D position of each individual cloud in heliocentric Galactic Cartesian Coordinates ($x$,$y$,$z$) using the central sky coordinates of the cloud ($l_{cen}$, $b_{cen}$) and the upper limit on the distance measured in \citet{redfield2008}, which broadly agrees with the CLIC morphology depicted in Figure~11 of \citet{frisch2011}. The LIC represents a special exception since it is thought that the solar system is surrounded by the LIC, though we are near its edge \citep{redfield2000}. The adopted distance for the LIC, $d=1.1 \; \rm pc$, is the distance to the center of the cloud based on detailed modeling of its 3D structure as described in \citet{linsky2019}.

In Table \ref{tab:clic6d}, we summarize the 3D heliocentric Galactic Cartesian positions $(x,y,z)$ and their associated velocities in the LSR frame, $(u,v,w)$, along with the underlying observational constraints from \citet{redfield2008} used to derive them. We additionally show the magnitude of the heliocentric Galactic Cartesian velocity vector with respect to the LSR ($v_{mag}$) and its associated uncertainty, $\sigma v_{mag}$. To compute $\sigma v_{mag}$, we randomly sample in ($V_0, l_0, b_0$) given the uncertainties on each parameter reported in Table \ref{tab:clic6d}. For each sample, we transform to ($u$,$v$,$w$) space, compute the magnitude of the velocity vector, and then take the standard deviation of all the samples to obtain the $\sigma v_{mag}$ values reported in Column (15) of Table \ref{tab:clic6d}. In \S \ref{sec:SN}, we will use $v_{mag}$ and $\sigma v_{mag}$ to model the evolution of the CLIC as driven by the most recent supernova shockwave that occurred in the Sco-Cen OB association $\rm \approx 1-2 \; Myr$ ago. 

\section{Evaluating the Potential Origin for the CLIC in Nearby Str\"{o}mgren Spheres} \label{sec:stromgren}

\citet{linsky2021} first proposed a possible interaction between the CLIC and nearby Str\"{o}mgren spheres, suggesting that the Local Bubble is fully ionized by the extreme ultraviolet (EUV) radiation from the $\rm B$ star $\rm \epsilon \; CMa$ which they argued produces a Str\"{o}mgren sphere with a Str\"{o}mgren radius $R_s \approx 160 \; \rm pc$ at a distance of $d=124 \; \rm pc$. If the surface of the Str\"{o}mgren sphere from e.g., $\rm \epsilon \; CMa$ interacted with recent supernova shockwaves from the Sco-Cen OB association, the process could lead to gas compression, cooling, and hydrogen recombination in the form of the CLIC, as observed elsewhere in the Galaxy \citep[e.g., toward the Cygnus Loop;][]{Raymond2020}. 

To elucidate this scenario, we need to know the 3D distribution and kinematics of the ionizing sources, the sizes of their corresponding Str\"{o}mgren spheres, and their trajectories in the past with respect to the CLIC constraints presented in \S \ref{sec:clicprops}. In \S \ref{subsec:strom_sample} we compile a sample of nearby ionizing sources, calculate their Str\"{o}mgren radii, determine their 6D phase information (3D positions and 3D velocities) and use their 6D phase information to compute their backward trajectories. In \S \ref{subsec:stromgren_viz} we compare the configuration of the most prominent Str\"{o}mgren sphere alongside the CLIC and a model for the evolution of the Local Bubble at key time snapshots over the past $\rm \approx 12 \; Myr$ to disfavor a purely Str\"{o}mgren sphere origin for the CLIC. 

\subsection{3D Position, Velocities, and Sizes of Nearby Str\"{o}mgren Spheres} \label{subsec:strom_sample}

Building on the work of \citet{linsky2021}, we compile a list of nearby stars that are either bright at EUV wavelengths or are known to be a hot white dwarf. For the bright EUV stars we adopt the sample of \citet{vallerga1998} and for the hot white dwarfs we adopt the sample of \citet{tat1999}. For the bright EUV stars, we place particular emphasis on characterizing the Str\"{o}mgren spheres around the B stars $\epsilon \; \rm CMa$ and $\beta \; \rm CMa$, as \citet{linsky2021} predicted these two stars should have by far the largest Str\"{o}mgren spheres in the solar vicinity.  Full details on the derivation of the Str\"{o}mgren radii, including assumptions on stellar radius, stellar temperature, and the local electron density are summarized in Appendix \ref{sec:stromgren_appendix}. 

In Appendix \ref{sec:stromgren_appendix}, we show the derived Str\"{o}mgren radius, $R_s$, for all stars in the sample with significant Str\"{o}mgren radii ($R_s > 5 \; \rm pc$) and whose full 3D space motions are constrained. The two hot B stars, $\epsilon$ and $\beta$ CMa have the largest Str\"{o}mgren radii, both significantly larger than 100\,pc. The three white dwarfs (WD2211-495, WD0232+035, and WD1056+516) with the largest Str\"{o}mgren spheres in the solar vicinity have radii of $R_s \approx $ 25\,pc. 

For each star in the sample we compute its heliocentric Galactic Cartesian coordinates ($x$,$y$,$z$) and its corresponding 3D space motions ($u$,$v$,$w$) with respect to the LSR, also shown in Appendix \ref{sec:stromgren_appendix}. To obtain $(x, y, z, u, v, w)$, we use a combination of astrometry (parallaxes and proper motions) and spectroscopy (radial velocity) data from Gaia DR3 and Hipparcos, supplemented by the broader literature when necessary. The full table of $(x, y, z, u, v, w)$ values for stars in the sample with significant Str\"{o}mgren radii are likewise summarized in Appendix \ref{sec:stromgren_appendix}. 

In addition to computing the $(x, y, z, u, w, w)$ data for each ionizing source, we leverage the 6D phase information to derive their trajectories over the past 20 Myr. We perform the dynamical tracebacks using the \texttt{galpy} package \citep{galpy}, which supports orbit integrations in a Milky-Way-like potential, for which we adopt the standard \texttt{MWPotential2014}, consisting of a spherical bulge and dark matter halo component, as well as a Miyamoto-Nagai disk component \citep{Miyamoto_1975}. We compute the dynamical tracebacks (from $ t=-20 \; \rm Myr$ to $t=0 \; \rm Myr$) for $\epsilon$ CMa and $\beta$ CMa. In Appendix \ref{sec:stromgren_orbit_appendix}, we show a 3D visualization of the orbital histories of all stars with Str\"{o}mgren spheres from Appendix \ref{sec:stromgren_appendix} over the past 3 Myr. 

\begin{figure*}[ht!]
\includegraphics[width=1.\textwidth]{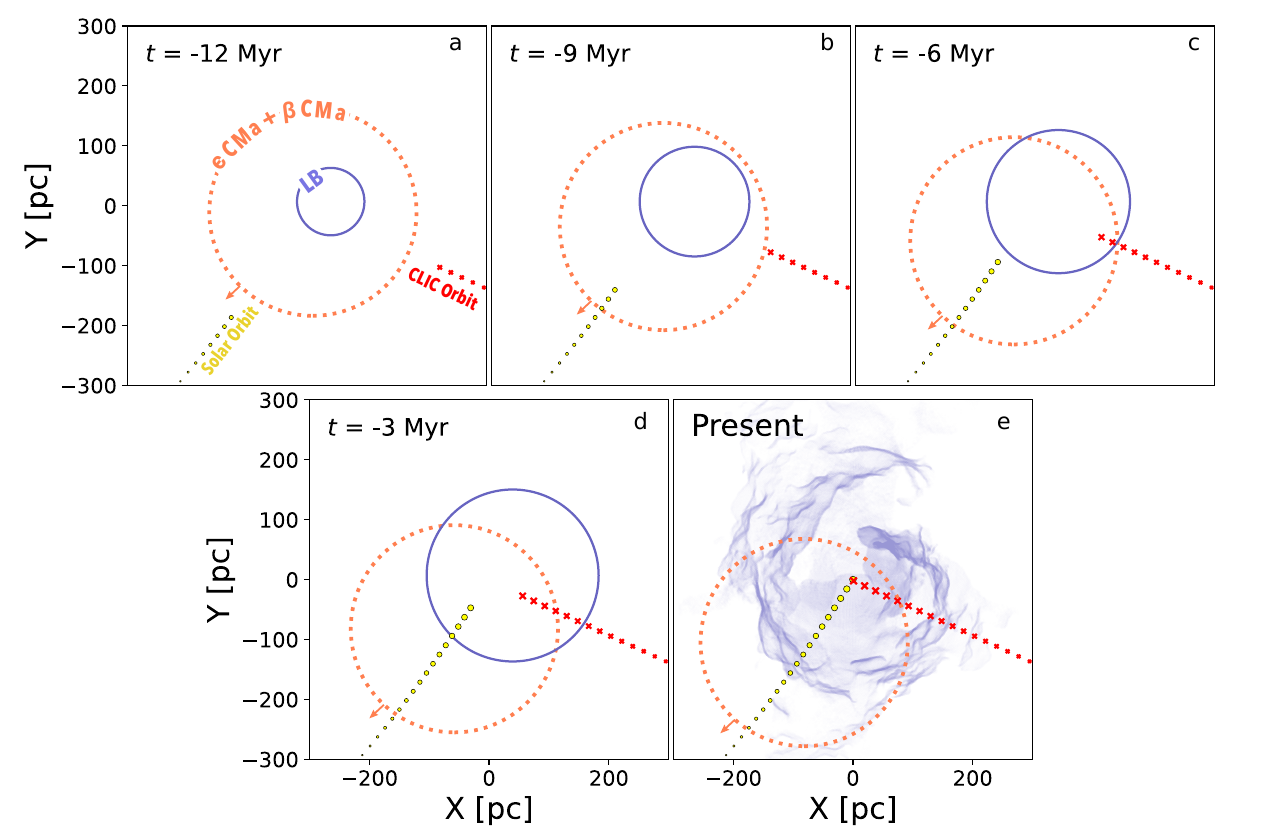}
\caption{Evolution of the combined $\epsilon$ CMa and $\beta$ CMa Str\"{o}mgren sphere (dotted orange ring), the Local Bubble (``LB", purple ring), the CLIC (average trajectory shown in red), and the Sun (trajectory shown in yellow). The Local Bubble at $t < 0 \; \rm Myr$ is modeled as an idealized, expanding spherical shell \citep{zucker2022}, with the more structured, non-spherical, boundary constrained by the present day 3D dust distribution \citep{ONeill_2024} shown at $t=0 \; \rm Myr$. \textit{Under the strong assumption of a constant electron density}, the Local Bubble was born within and has partially overlapped with the boundary of a Str\"{o}mgren sphere over its lifetime. Tracing back its present day trajectory far beyond its presumed lifetime, we find that the CLIC has been traveling within the combined $\beta$ CMa and $\epsilon$ CMa Str\"{o}mgren sphere for the past several million years, whose trajectory is nearly perpendicular to the trajectory of the CLIC. The combination of the CLIC's position interior to the Str\"{o}mgren sphere and its motion transverse to the CLIC disfavors an origin for the CLIC tied solely to this prominent nearby Str\"{o}mgren sphere . \label{fig:stromgren_evolution}}
\end{figure*}

\subsection{History of Nearby Str\"{o}mgren Spheres, the Local Bubble, and the CLIC} \label{subsec:stromgren_viz}

The two hot B stars, $\epsilon$ and $\beta$ CMa, have by far the largest Str\"{o}mgren radii in the solar vicinity, with $R_s = 151 \; \rm pc$ and $R_s = 120 \; \rm pc$, respectively. The star $\epsilon$ CMa has an age of 22.5 Myr \citep{tetzlaff_2011}, and passed within 10 pc of the Sun between $4-5$ Myr ago \citep{Shull_2025}. The star $\beta$ CMa has an age of 12.4 Myr \citep{Mazumdar2006}. Tracing back the trajectories of these two stars over the past 12 Myr, we also find that they share similar trajectories, lying on average 30 pc apart over the past 6 Myr, and passing roughly 16 pc from each other at closest approach 3 Myr ago. In Appendix \ref{sec:stromgren_coevolve}, we show the evolution of $\epsilon$ CMa and $\beta$ CMa and their respective Str\"{o}mgren spheres over the past 12 Myr, assuming the Str\"{o}mgren spheres evolve completely independently. However, given the close separation of the two stars, their two Str\"{o}mgren will interact to produce a larger, combined Str\"{o}mgren sphere with a radius $R_s = 173 \; \rm pc$ (see full details in Appendix \ref{sec:stromgren_coevolve}). Accordingly, in this section, we model the combined effects of a single Str\"{o}mgren sphere stemming from $\epsilon$ CMa and $\beta$ CMa assuming an average trajectory of the two stars. 

In Figure \ref{fig:stromgren_evolution}, we show the combined $\epsilon$ CMa and $\beta$ CMa Str\"{o}mgren sphere, the evolution of the Local Bubble's shell \citep{zucker2022}, the average position of the CLIC clouds (derived from Table \ref{tab:clic6d} assuming purely linear backward trajectories\footnote{Unlike the stars, we do not trace back the trajectory of the gaseous CLIC using \texttt{galpy}. However, including the gravitational acceleration from the Milky Way, as \texttt{galpy} does, would not lead to a significant change in the cloud trajectories.}), and the Sun's position at five time snapshots over the past twelve million years ($t = -12, -9, -6, -3$ and $0$ Myr). 

According to \citet{zucker2022}, the present-day Local Bubble began forming about 14 Myr ago due to a series of supernovae that started exploding in the Upper Centaurus Lupus and Lower Centaurus Crux (LCC) sub-clusters in the Sco-Cen OB association. \citet{zucker2022} model the evolution of the Local Bubble's dense shell given the 3D spatial, dynamical, and age constraints of young stellar clusters sequentially forming on its expanding surface over time (purple rings in Figure \ref{fig:stromgren_evolution}). 

\textit{Under the strong assumption of a fixed electron density in the local vicinity} --- implying that the Str\"{o}mgren radius has been roughly constant --- the Local Bubble was born within and has at least partially overlapped with the boundary of the $\epsilon$ CMa and $\beta$ CMa Str\"{o}mgren sphere over its lifetime. Clearly the electron density in the solar vicinity has not been constant over the past $\approx 12$ Myr, as we assume in Figure \ref{fig:stromgren_evolution}. However, because $\epsilon$ CMa and $\beta$ CMa have been located close to the epicenter of the supernova explosions driving the bubble's expansion, assuming a higher electron density in the past would not significantly alter this picture. The Str\"{o}mgren radius scales as the electron density $n_e^{-\frac{2}{3}}$. Assuming a $3\times$ higher electron density would only decrease the Str\"{o}mgren radii by roughly a factor of two, and the Local Bubble would have still at least partially overlapped with the Str\"{o}mgren sphere over much of its lifetime. As a result, the existence of the combined $\epsilon$ CMa and $\beta$ CMa Str\"{o}mgren sphere should be considered in future models of the Local Bubble's evolution, as well as characterization of its thermal and ionization state. 

The properties and motions of nearby Str\"{o}mgren spheres also carry potential implications for the origin of the CLIC. First, as seen in Figure \ref{fig:stromgren_evolution}, the CLIC entered the combined Str\"{o}mgren sphere (orange dotted ring) roughly 9 Myr ago, and has been traveling interior to the sphere surface since then. And second, the trajectory of the CLIC is roughly transverse to the trajectory of the Str\"{o}mgren sphere, counter to the idea that the largest nearby sphere could account for the CLIC's current 3D space motion. The combination of these two factors disfavors, but does not disqualify (or rule out) a purely Str\"{o}mgren sphere origin for the CLIC. However, if the electron density was different in the past, the trajectory of the CLIC, rather than being interior to the Str\"{o}mgren sphere, may have been tangent to the Str\"{o}mgren sphere surface. If supernova shockwaves stemming from the Sco-Cen association propagated into the Str\"{o}mgren sphere surface --- a region of lower magnetic field strength --- this propagation could have lead to more gas compression, gas cooling, and increased hydrogen recombination, potentially making it easier for the CLIC to form \citep[see further discussion in][]{linsky2021}. We explore the possibility of a supernova origin in \S \ref{sec:SN} and emphasize that depending on the electron density, the Str\"{o}mgren sphere either played no role in the CLIC's evolution or could have aided the formation of the CLIC in the supernova scenario, depending on the exact orientation between the supernova shockwaves and the Str\"{o}mgren sphere surface over time (see also discussion of potential interactions with smaller Str\"{o}mgren spheres in Appendix \ref{sec:stromgren_orbit_appendix}). 

\section{A Supernova Origin for the CLIC} \label{sec:SN}
Having determined that a Str\"{o}mgren sphere alone likely could not have accounted for the formation of the CLIC, in this section we expound upon the possibilities of a supernova-driven origin. In \S \ref{subsec: SN_setup}, we summarize the growing body of evidence that the Local Bubble was formed by a series of supernovae stemming from the Sco-Cen OB association, and propose that the most recent supernova that occurred within its UCL subcluster may explain the observed spatial and velocity distribution of the CLIC. In \S \ref{subsec:modeling}, we explore this possibility in more detail, presenting the formalism for the radial and velocity evolution of a single supernova remnant in a uniform ambient medium. We describe how the evolution of this supernova remnant can be constrained by the present day velocities of the CLIC --- subject to some deceleration (\S \ref{subsec:deceleration}) --- under the assumption that individual CLIC clouds formed in the expanding shell as it fragmented and condensed due to tenuous material being swept up inside the pre-evacuated cavity of the Local Bubble. We then fit for the parameters governing the shell's evolution in \S \ref{subsec:dynestyfitting}. 

\subsection{Supernovae and the Local Bubble} \label{subsec: SN_setup}

As alluded to in \S \ref{subsec:stromgren_viz}, several lines of evidence suggest that a series of supernova explosions beginning $\approx 10-15$ Myr ago was responsible for carving out the Local Bubble. First, studies place the Sco-Cen OB association at the center of the Bubble when it first started forming, and this OB association is expected to have produced many supernovae in the recent past \citep{maizapellaniz2001,fuchs2006,breitschwerdt2016}. Second, the present-day properties of the Local Bubble's shell (e.g., its size and momentum) are consistent with the number of supernovae predicted to have exploded in Sco-Cen. And third, signatures of $\rm ^{60}Fe$ in deep-sea crusts --- a radioactive isotope produced predominantly in supernova explosions  --- indicate that multiple supernovae occurred in the solar neighborhood over the past $\rm \approx 10 \; Myr$ and have polluted the Sun's local interstellar environment  with the nucleosynthetic products of these explosion  \citep{wallner2021, feige2012}.

\citet{maizapellaniz2001} first proposed the Sco-Cen OB association as the progenitor association for the Local Bubble, using evolutionary synthesis modeling to argue that its constituent clusters UCL and LCC must have produced $\approx 20$ supernova explosions in the past 10-12 Myr. Tracing back the motions of UCL and LCC over the past $5-7$ Myr using data from the Hipparcos mission, \citet{maizapellaniz2001} found that UCL and LCC were closer to the present day position of the Sun (and the center of the Local Bubble) in the past. Building on the study of \citet{maizapellaniz2001}, \citet{fuchs2006} computed the backward trajectories of all nearby B stars with Hipparcos and confirmed that UCL and LCC were the only clusters that could have powered the Local Bubble's expansion.  Fitting an Initial Mass Function (IMF) to the present day stellar cluster members, \citet{fuchs2006} argued that between $14-20$ supernovae have exploded in UCL and LCC, consistent with the estimate of \citet{maizapellaniz2001}. \citet{fuchs2006} also argued that this number of supernovae was sufficient to carve out the present day size of the Local Bubble's cavity. 

Using new spatial and dynamical constraints from Gaia, \citet{zucker2022} found that UCL and LCC met at the same location $\approx 15-16$ Myr ago, and would have been at the center of the Bubble when it first started forming. \citet{zucker2022} further found that all nearby molecular clouds lie on the surface of the Local Bubble.  Using the 3D space motions of these clouds' embedded stars to study the past trajectories of their young stellar clusters, \citet{zucker2022} found clear evidence for a global expansion of the Local Bubble, arguing that the Bubble's surface is currently expanding at a rate of $\rm 7 \; km \; s^{-1}$ and contains a total swept-up mass of $\approx 1.4$ million $\rm M_\sun$. Based on the amount of momentum injection required by supernovae to sweep up this shell mass given its present day expansion velocity, \citet{zucker2022} argued that $\approx$ 15 supernovae were required to have exploded in UCL and LCC to power the bubble's expansion, in strong agreement with estimates from evolutionary synthesis modeling. 

Of the $\approx 14-20$ supernovae predicted to powered the Bubble's expansion, \citet{breitschwerdt2016} estimated that the most recent explosion occurred in the UCL cluster $\approx$ 1.5 Myr ago. Incorporating new data from Gaia, \citet{neuhauser2020} found kinematic evidence that the runaway star $\zeta$ Oph and the radio pulsar PSR B1706-16 were released by a supernova in a binary that took place in UCL $1.78 \pm 0.21$ Myr ago, in agreement with the \citet{breitschwerdt2016} result based on IMF modeling. Based on the present day boundary of the Local Bubble defined via 3D dust maps \citep[][]{ONeill_2024}, UCL was located very close to the edge of the Local Bubble when the most recent supernova would have occurred. Therefore, tenuous material in the largely evacuated extant cavity could have been swept-up by the most recent supernova that occurred in UCL circa $\rm \approx 1-2\; Myr$ ago, potentially aided by interaction with the combined $\epsilon$ CMa and $\beta$ CMa Str\"{o}mgren sphere or a smaller Str\"{o}mgren sphere. 

\begin{figure}[ht!]
\includegraphics[width=0.5\textwidth]{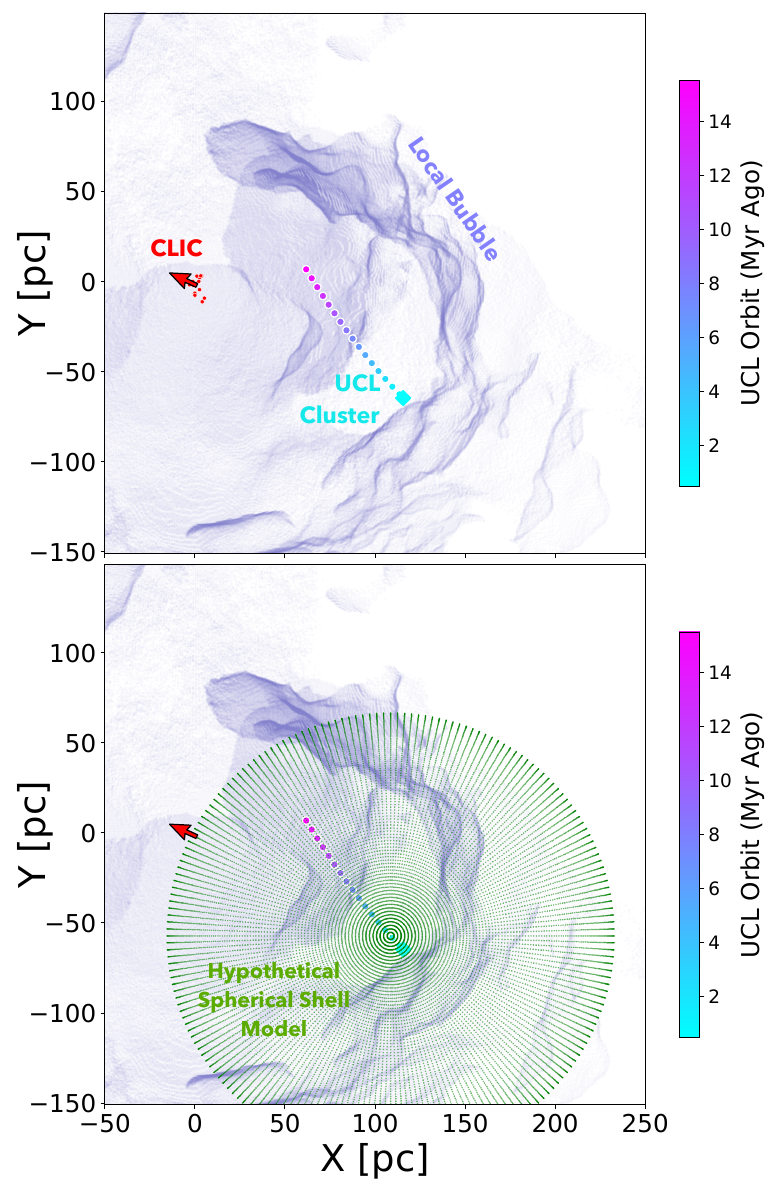}
\caption{\textit{Top}: Present day position of the UCL stellar cluster (cyan diamond) and the CLIC (red dots). The average 3D space motion of the CLIC (red arrow) is also shown. The colored path shows the trajectory of the UCL cluster from its birth 16 Myr ago to the present day, colored by time in the past. The current 3D space motions and 3D positions of the CLIC clouds with respect to UCL may be explained by a supernova that exploded in the cluster within the past $\rm \approx \; 1-2 \; Myr$, potentially sweeping up the tenuous material inside the pre-evacuated Local Bubble cavity and forming the CLIC. A present day model for the Local Bubble \citep[structured purple surface;][]{ONeill_2024} is constrained by the observed 3D distribution of dust \citep{Edenhofer2024}, and is the same model shown in the bottom right panel of Figure \ref{fig:stromgren_evolution}. \textit{Bottom}: We overlay a hypothetical model for a spherical shell remnant (green points) for a supernova that may have gone off $1.78 \; \rm Myr$ ago in UCL \citep[e.g.,][]{neuhauser2020}, adopting a shell radius that places the CLIC on its surface in the present day. The current space motion of the CLIC (red arrow) is nearly perpendicular to the surface of such a shell. 
\label{fig:UCL_CLIC_LB} }
\end{figure}

In the top panel of Figure \ref{fig:UCL_CLIC_LB}, we show the trajectory of the UCL cluster since its birth circa 16 million years ago (colored by time in the past). We adopt the mean 3D position and mean 3D motion of all stars in all subclusters classified as belonging to UCL in the recent census of \citet{Ratzenbock2022} (see their Table 3). We trace back the trajectory of the UCL cluster using \texttt{galpy}, as described for the stars with significant Str\"{o}mgren spheres in \S \ref{sec:stromgren}. Alongside the UCL cluster, we show a model for the present day boundary of the Local Bubble from \citet{ONeill_2024} \citep[based on the 3D dust map from][]{Edenhofer2024} and the current average motion of the CLIC. In the present day, UCL is $\rm \approx 130\; pc$ away from the average position of the CLIC. The UCL cluster is also $\rm \approx 10\; pc$ from the wall of the Local Bubble at its closest point. 

The Gaia era has ushered in a new model for the configuration of the Local Bubble with respect to Loop I, whose ``S1" subshell was argued to host the CLIC clouds in previous work \citep{frisch2011, frisch2017}.  Recent Gaia-based constraints \citep{panopoulou2021} placed sections of the near edge of Loop I at distances of $\rm 112 \pm 17 \; pc$ and $\rm 135 \; \pm 20 \; pc$, respectively, well beyond the current location of the CLIC at distances $\rm < 15 \; pc$ from the Sun. The revised distance estimate for Loop I places it at a consistent distance with the surface of the Local Bubble shown in Figure \ref{fig:UCL_CLIC_LB}. Thus, if Loop I and the Local Bubble are interacting \citep[as modeled in][]{breitschwerdt2000}, they are doing so at distances $\rm >100 \; pc$, and their interaction could not have given rise to the CLIC. 

Instead, we argue that, for the most recent supernova that occurred $\rm \approx 1-2 \; Myr$ ago in UCL, the shockwave traveling towards $+x$ (the Galactic center direction) would have propagated into a medium with typical densities of $\rm n \approx 100 \; cm^{-3}$ (corresponding to the Local Bubble's surface) while the shockwave traveling towards $-x$ (the Galactic anti-center direction) would have propagated into a tenuous medium with densities of $\rm n \approx 0.01 -  0.1 \; cm^{-3}$. 

Assuming a roughly spherical expansion of the remnant, we postulate that both the high average speed of the CLIC ($\rm \approx 26 \; km\; s^{-1}$ with respect to the LSR and $4\times$ faster than dense clouds on the Local Bubble's surface) as well as its direction could be explained by the propagation of this most recent supernova shockwave into the tenuous cavity of the Local Bubble towards the Galactic anti-center. In the bottom panel of Figure \ref{fig:UCL_CLIC_LB} we overlay a hypothetical model for a spherical shell remnant (shown in green), for a supernova that went off $1.78 \; \rm Myr$ ago \citep[e.g.,][]{neuhauser2020} in UCL with a radius that places the CLIC in the shell of such a remnant in the present day. As apparent in Figure \ref{fig:UCL_CLIC_LB}, the mean velocity vector of the CLIC is nearly perpendicular to the surface of such a shell. Computing the angle between the 3D normal vector to the shell's surface at the CLIC's 3D position and the CLIC's 3D space motion vector, we find a difference in angle of only $8^\circ$. When projected onto the XY plane, as shown in Figure \ref{fig:UCL_CLIC_LB}, the difference in angle is only $2^\circ$.

In \S \ref{subsec:modeling} we explore this physical scenario in more detail by modeling the evolution of such a remnant, given the trajectory of UCL, the possible range of explosion times, and the 3D positions and 3D space motions of clouds in the CLIC. 

\begin{figure*}[ht!]
\includegraphics[width=1.\textwidth]{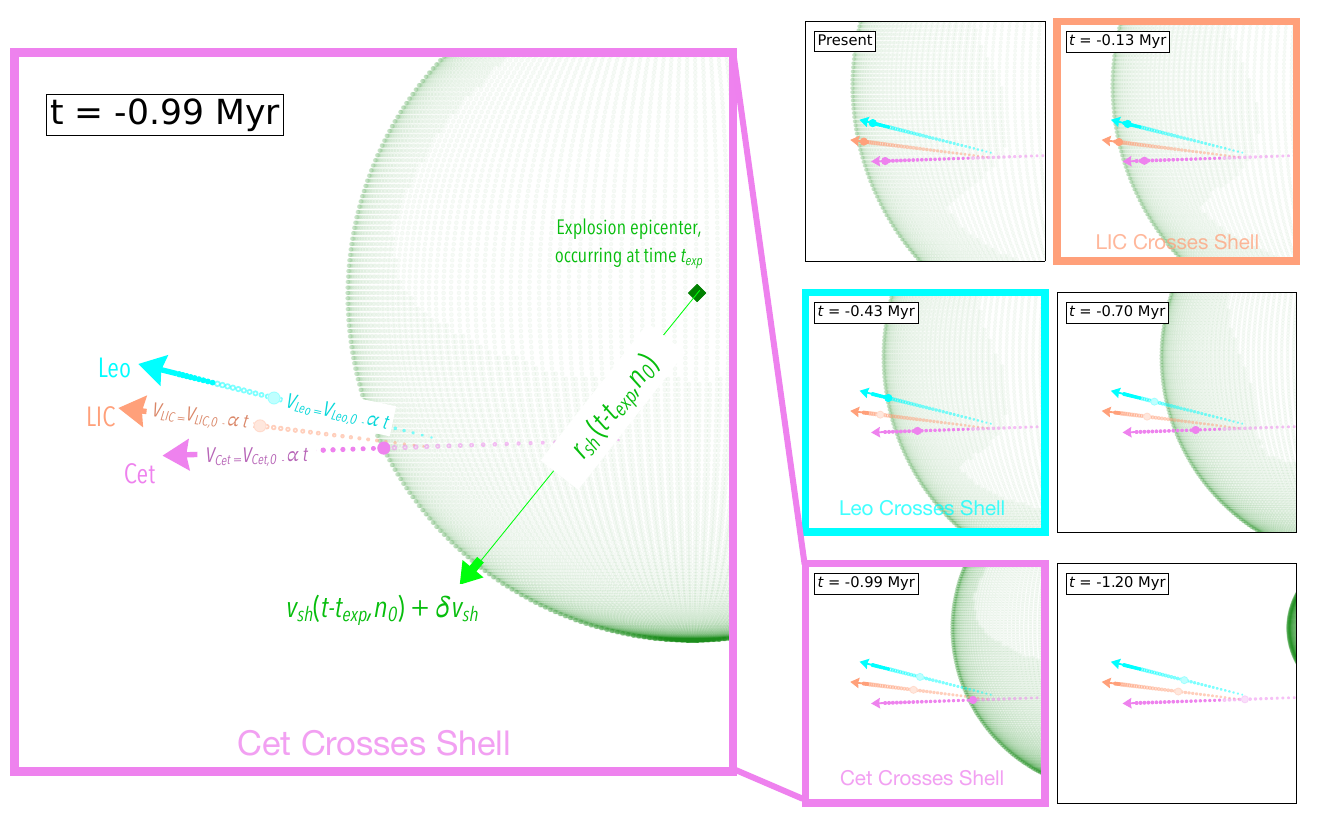}
\caption{A pictorial explanation of the likelihood function used to fit for the evolution of the supernova shell. We assume each cloud forms in the shell at the time of crossing. Three clouds --- Cet (pink trace), LIC (salmon trace), and Leo (cyan trace) --- are shown here as examples, but the full likelihood is fit over the entire ensemble of clouds. The present day 3D positions and 3D space motions of the clouds (marked with the pink, salmon, and cyan vectors) define the backward trajectory of the clouds, subject to a deceleration term parameterized by $\alpha$. Given a model for the radial and velocity evolution of the supernova shell (defined by the ambient volume density of hydrogen nuclei $n_0$) with the epicenter of the expansion defined by the explosion time $t_{exp}$, each cloud will first intersect the shell at a single point in time, as shown in the time snapshots in the sub-panels at right. For this example, LIC crosses the shell at $t\approx-0.13 \; \rm Myr$ (sub-panel framed in salmon), Leo at $ t\approx-0.43 \; \rm Myr$ (sub-panel framed in cyan) and Cet at $t\approx-0.99 \; \rm Myr$ (sub-panel framed in pink). Our likelihood function seeks to minimize the velocity difference between the clouds and the shell at their respective times of intersection, modulo a small velocity shift $\delta v_{sh}$, which can account for additional sources of uncertainty on the shell velocity due to the simplifications of our model. \label{fig:likelihood}}
\end{figure*}

\subsection{Modeling the Shell Evolution: Formalism for Expansion of a Supernova Remnant} \label{subsec:modeling}
The 3D velocity vectors for the CLIC clouds in Table \ref{tab:clic6d} have similar directions, oriented radially away from where the center of the UCL cluster would have been circa $1-2$ Myr ago (bottom panel of Figure \ref{fig:UCL_CLIC_LB}). However, the clouds exhibit a wide range of velocity amplitudes,  traveling anywhere from $\rm 10 - 48 \; km\; s^{-1}$ with respect to the LSR (see Column 14 in Table \ref{tab:clic6d}). This broad range of speeds suggests that while the clouds may have a common origin, they did not form instantaneously together. We posit that the range of velocity amplitudes observed for the CLIC may be explained by the clouds' having been swept up at different periods in the shell's evolution during its pressure-driven snowplow phase, potentially caused by the Vishniac instability \citep[e.g.,][]{Vishniac_1983}, which can result in mass clumping on the shell's surface. If this theory is true, the velocities of the CLIC clouds can provide a constraint on the evolution of the shell as a function of time. It is also possible that the CLIC clouds pre-existed and were simply accelerated (rather than swept up and formed) by the supernova explosion, but we do not explore that scenario in this work. 

The expansion of a single supernova remnnant into a uniform medium has been well-studied for decades via spherically symmetric models. The evolution of these supernova remnants can be characterized by several phases, namely the ejecta-dominated free expansion phase, an energy-conserving Sedov-Taylor phase, the time of shell formation when radiative losses become important, the pressure-driven snowplow phase, and the final momentum-conserving phase --- concluding when the remnant eventually merges with the surrounding interstellar medium. In each stage, the power-law expansion of the supernova remnant's radial evolution $r$ as a function of time $t$ since the explosion can be approximated as $r \propto t^\eta$, where $\eta = 1$ for free expansion, $\eta = 2/5$ for Sedov-Taylor, $\eta = 2/7$ for the pressure-driven snowplow, and $\eta = 1/4$ for the momentum-conserving snowplow phase. 

To model the evolution of the most recent supernova explosion from UCL in a largely evacuated cavity, we find that the pressure-driven snowplow stage ($\eta = 2/7$) is the dominant phase of evolution over the temporal timescales targeted in this work. Assuming a fixed energy injection per supernova $E_{SN} = 10^{51} \; \rm erg \; s^{-1}$,  \citet{Kim_Ostriker_2015} analytically characterized the evolution in the radius, post-shock temperature, swept-up mass, and momentum of a single supernova remnant as a function of the volume density of hydrogen nuclei in the ambient interstellar medium, $n_0$.  They find that the time of shell formation $t_{sf}$ (marking the beginning of the pressure-driven snowplow phase) is parameterized as
\begin{equation} \label{eq:shell_formation}
t_{sf} = 4.4 \times 10^{4} \; {\rm yr} \; \bigg(\frac{n_0}{1 \; \rm cm^{-3}}\bigg)^{-0.55}.
\end{equation}

Assuming that the present day average density of the Local Bubble ($n \approx 0.01 \; \rm cm^{-3}$) is a lower limit on the density, $n_0$, of the bubble circa 2 million years ago (a valid assumption given that one additional supernova has gone off in the recent vicinity in UCL since then), the longest plausible time for shell formation to occur would be $\approx 600,000$ years, or less than one third of the likely age of the most recent supernova explosion occurring in the UCL association.  For densities between $n_0 \approx$ 0.05--0.1\;cm$^{-3}$, shell formation occurs between 150,000--230,000 years. 

Accordingly, if the pressure-driven snowplow phase dominates the bubble's evolution, the radius of the supernova shell $r_{sh}$ follows a power-law with $\eta = 2/7$, parameterized as a function of both the time of shell formation and the radius of the bubble at the time of shell formation
\begin{equation} \label{eq:shell_radiuss}
r_{sh}(t) = r_{sf} \times \bigg(\frac{t - t_{exp}}{t_{sf}}\bigg)^{2/7},
\end{equation}
where $r_{sf} = 22 \; {\rm pc} \times \big(\frac{n_0}{1 \; {\rm cm}^{-3}}\big)^{-0.42}$. The term $t_{exp}$ refers to the time of the most recent supernova explosion in UCL, with $t=0$ corresponding to the present day (0 Myr) and becoming more negative as $t$ approaches $t_{exp}$. The time of explosion additionally sets the epicenter of the explosion, corresponding to the average 3D position of the UCL cluster in its past trajectory when the supernova went off (see Figure \ref{fig:UCL_CLIC_LB}). 

The velocity evolution of the shell then follows as 
\begin{equation} \label{eq:shell_velocity}
v_{sh}(t) = \frac{2}{7} \times \bigg(\frac{r_{sf}}{t_{sf}}\bigg) \times \bigg(\frac{t - t_{exp}}{t_{sf}}\bigg)^{-5/7}.
\end{equation}

\noindent Since this model for the velocity evolution is highly idealized, we include an additional term, $\delta v_{sh}$, that will provide our fit additional freedom to account for systematic uncertainties on our modeled shell speed. 

The free parameters governing the shell thus include the ambient volume density of hydrogen nuclei $n_0$, the explosion time $t_{exp}$, and the turbulent shell velocity term $\delta v_{sh}$, which models small offsets between the shell and the ensemble of clouds at the time of intersection given our simplified model for the shell's evolution. To fit for the velocity evolution of the shell, we adopt a Gaussian log-likelihood of the following form
\begin{equation} \label{eq:logl}
\begin{split}
\log(L)=& - \frac{1}{2}   \sum_i \Bigg( \frac{[v_{sh}(n_0, t_i)+\delta v_{sh}-v_{cl,i}(\alpha,t_i)]^{2}}{\sigma_{v_{cl,i}}(f)^2} \\
&+ \log(2\pi \sigma_{v_{cl,i}}(f)^{2})  \Bigg).
\end{split}
\end{equation}

\noindent Here, $v_{sh}(n_0, t_{i})$ is the velocity of the expanding shell governed by Equation \ref{eq:shell_velocity}.  The $v_{sh}(n_0, t_{i})$ term is evaluated at $t_i$, corresponding to the time when the ith cloud first intersects the shell. The term $v_{cl,i}(\alpha,t_i)$ is the velocity that the $ith$ cloud has at this time of intersection. Rather than assume that the present day velocity of each cloud is constant as a function of time, we allow the clouds to decelerate dependent on a free parameter $\alpha$, which is based on the mean present day velocity of the clouds and the ratio of the cloud density to the ambient interstellar density, as discussed in \S \ref{subsec:deceleration}.

The term $\sigma_{v_{cl,i}}$ is the uncertainty in the $ith$ cloud's velocity at the time of intersection.  While we allow the magnitude of the cloud velocity to change, we assume the uncertainty on the magnitude of the cloud velocity remains constant as a function of time but include an additional free parameter, $f$, to account for additional uncertainties tied to the evolution of the cloud velocities over time. We assume that $\sigma_{v_{cl,i}}(f)$ = $\sigma v_{mag_{LSR},i} + (f\times \sigma v_{mag_{LSR},i})$, where $\sigma v_{mag_{LSR},i}$ is the present day uncertainty on the measured magnitude of the $ith$ cloud's velocity (see Column 15 of Table \ref{tab:clic6d}). We emphasize that $f$ accounts for the uncertainty on the cloud velocities \textit{in the past}, rather than the present day, and we have simply made a choice to scale this uncertainty as a function of their present day values.  A pictoral representation of our likelihood function is shown in Figure \ref{fig:likelihood}. In our simple model, each cloud is assumed to inherit the velocity of the expanding shell at the time of its birth, before decoupling and evolving independently from both the shell and the other clouds forming within it. This assumption is discussed further in the context of potential caveats of our analysis in \S \ref{subsec:uncertainties}.

\subsection{Modeling the Shell Evolution: Formalism for Cloud Deceleration} \label{subsec:deceleration}

In order to model the decrease in the magnitude of the clouds' velocities as a function of time since their presumed formation in the supernova shell, we assume that the deceleration in the velocity is dominated by ram pressure. Given densities of a few hundredths of a particle per cubic centimeter, and temperatures anywhere between $10^4 - 10^6 \; \rm K$ interior to the bubble (see \S \ref{sec:intro}) the collisional timescale is anywhere from a few dozen to a few hundred years, which, while long, is much less than the million year timescales considered in our modeling, so invoking ram pressure is not an unreasonable assumption. The ram pressure $P$ which acts on a cloud moving through the interstellar medium depends on the ambient mass density $\rho_0$ and the relative velocity between the cloud ($v_{cl}$) and the ambient interstellar medium. Given the very high present day mean cloud velocity of $\rm 26 \; km \;s ^{-1}$ (see Table \ref{tab:clic6d}), we approximate the ram pressure as 
\begin{equation} \label{eq:ram}
P \approx \rho_0 \times v_{cl}^{2}.
\end{equation}

\noindent This results in $v_{cl}$ evolving with time as follows:

\begin{equation} \label{eq:cloudvel}
v_{cl}(\gamma, t) = \frac{v_{cl,0}}{1+v_{cl,0} \gamma t} ,
\end{equation}
where $v_{cl,0}$ is the cloud's velocity in the present day ($t=0$; see Column 14 of Table \ref{tab:clic6d}), with the cloud velocity $v_{cl}$ increasing in magnitude as $t$ becomes more negative, approaching $t_{exp}$. The term $\gamma$ depends on the ratio of the ambient mass density to the cloud surface density. Assuming a spherical cloud geometry
\begin{equation} 
\gamma = \frac{\rho_0}{\Sigma_{cl}} = \frac{3}{4} \times \frac{1}{r_{cl}} \times \frac{\rho_{0}}{\rho_{cl}},
\end{equation}
with $r_{cl}$ being the cloud radius, $\rho_{cl}$ the cloud mass density, and $\rho_{0}$ the ambient mass density of the interstellar medium into which the clouds are decelerating.

Expanding Equation \ref{eq:cloudvel} in the $v_{cl,0} \, \gamma \, t$ term leads to the following linear approximation for the change in velocity
\begin{equation} \label{eq:simple_cloudvel}
v_{cl} (\alpha, t) = v_{cl,0} - \alpha \times t,
\end{equation}
where $\alpha = v_{cl,0}^{2} \times \gamma$. Note that the influence of the velocity field of the ambient interstellar medium can always be absorbed in $\alpha$. We adopt the linearized formula for the cloud velocities given by Eq. \ref{eq:simple_cloudvel} (parameterized by $\alpha$) over the formula for the cloud velocities given by Eq. \ref{eq:cloudvel} (parameterized by $\gamma$) for computational reasons. However, we have confirmed that our key results (regarding the ambient density and time of supernova explosion) do not change if we adopt the non-linearized formula for the cloud velocities (given by Eq. \ref{eq:cloudvel}), as well as if we sample for $\gamma$ instead of $\alpha$. We have also confirmed that our results are robust to the exact parameterization of how we model the error on the cloud velocities in the past. Specifically, rather than scaling the present day cloud velocities by $f$, we also tested adding an additional error term in quadrature with the present day cloud velocities, and again find consistent results for the ambient density and time of supernova explosion. We fit for $\alpha$, along with the other parameters governing the shell's evolution in \S \ref{subsec:dynestyfitting}.

\subsection{Fitting the Shell Evolution} \label{subsec:dynestyfitting}

To fit for the parameters governing the shell's evolution and the deceleration of the CLIC, we use the Nested Sampling code \texttt{dynesty} \citep{Speagle2020}. We adopt the log-likelihood function given in Equation \ref{eq:logl} and priors informed by extant studies. 

For $n_0$ (the ambient density of hydrogen nuclei at the time of the most recent supernova in UCL) we adopt a truncated log-normal prior with a mean of $\rm 0.16 \; cm^{-3} \rm $ and a standard deviation of a factor of three, with a minimum density of $\rm 0.01 \; cm^{-3}$ and a maximum density of $\rm 2.7 \; cm^{-3}$. Note that the factor of three describes the spread in the logarithmic domain and is not the same as the standard deviation of the log-normal distribution in linear space. The mean prior density of $n_0 = 0.16 \; \rm cm^{-3}$ is obtained by taking the median present day density of all material inside the Local Bubble within 100 pc of the midplane, where the volume density distribution has been derived from 3D dust maps \citep{Leike2020}\footnote{The conversion from differential extinction $s_x$ to volume density of hydrogen nuclei $n_{\rm H}$ was adopted from \citet{Zucker_2021}: $n_{\rm H} = 880 \; {\rm cm}^{-3} \times s_x$ (see their \S 2.1.1).} and the model for the Local Bubble's surface taken from \citet{ONeill_2024}. The lower limit on $n_0$ is set by the present day inferred density of the bubble of $\rm 0.01 \; cm^{-3}$ just beyond the CLIC \citep{frisch2007}, while the upper limit of $\rm 2.7 \; cm^{-3}$ is the inferred density of the bubble 14 Myr ago before any supernova went off \citep[based on the results of][]{zucker2022}.

For the time of the most recent supernova explosion in the UCL, $t_{exp}$, we adopt a truncated normal prior with a mean of $\rm -1.78 \; Myr$ and a standard deviation of $\rm 0.21\; Myr$ (lower and upper bounds of $-0.5$ and $\rm -3 \; Myr$) based on the results of \citet{neuhauser2020}.

For the $\delta v_{sh}$ parameter, which accounts for small velocity offsets between the shell velocity and the ensemble of cloud velocities due to our simplified model, we adopt a truncated normal prior with a mean of $\rm 0 \; km \; s^{-1}$ and a standard deviation of $\rm 5 \; km \; s^{-1}$, with a lower and upper bound of $\rm \pm 20 \; km \; s^{-1}$. 

For the $\alpha$ parameter, which models the clouds' deceleration, we adopt a truncated normal prior with a mean of $\rm \; 20 \; pc \; Myr^{-2}$, a standard deviation of $\rm \; 5 \; pc \; Myr^{-2}$, and a generous lower and upper bound of $[\rm 0,  100 \; pc \; Myr^{-2}]$. As detailed in \S \ref{subsec:deceleration}, $\alpha = v_{cl,0}^{2} \times \gamma$, with $\gamma = \frac{\rho_0}{\Sigma_{cl}} = \frac{3}{4} \times \frac{1}{r_{cl}} \times \frac{\rho_{0}}{\rho_{cl}}$. Therefore, assuming a mean cloud radius of $r_{cl} = 2.5 \; \rm pc$ \citep{redfield2008}, a mean mass density ratio between the CLIC and the ambient interstellar medium of $\frac{\rho_{cl}}{\rho_0} = 10$ \citep{Linsky_2022}, and a mean present day cloud velocity of $v_{cl,0} = 26 \; \rm km \;s ^{-1}$ (see Table \ref{tab:clic6d}) results in a mean value of $\alpha = 20 \rm \; pc \; Myr^{-2}$.

Finally, to ensure that the parameter $f$ --- governing the fractional amount that the formal uncertainty on the cloud's past velocity is underestimated --- is never negative, we sample in the logarithm of $f$ instead \citep[see e.g. the tutorial in][]{emcee_2013}. We place a flat prior on $\log(f)$ over the range $-2.3$ to $2.3$, allowing the cloud's past velocity uncertainty to be underestimated from roughly 10\% all the way up to a factor of 10. However, we also tested whether sampling for $f$ linearly over the range 0 to 10 affected our results (allowing for no inflation in the uncertainties in the lower bound of zero), and find consistent results across all model parameters, irregardless of sampling in $\log(f)$ or $f$.

\begin{figure}[ht!]
\includegraphics[width=0.5\textwidth]{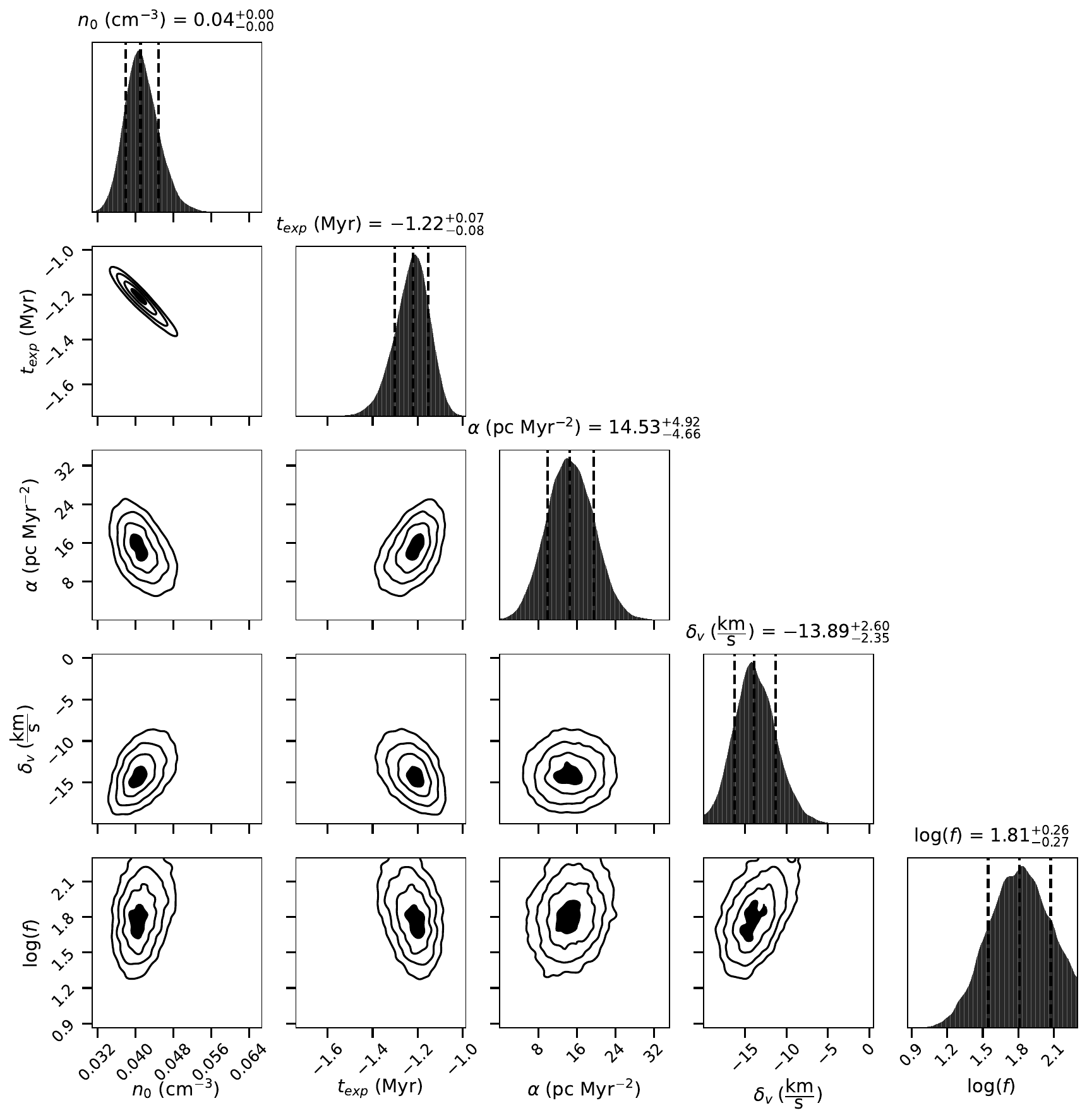}
\caption{1D and 2D marginal distributions (``corner plot") of the model parameters governing the evolution of the most recent supernova that exploded in UCL and potentially drove the formation of the CLIC. Parameters include the ambient density of the interstellar medium prior to the explosion ($n_0$), the time the supernova exploded in UCL ($t_{exp}$), a turbulent velocity parameter ($\delta v_{sh}$) that models small shifts between the shell velocity and the ensemble of cloud velocities at the time of intersection, a parameter $\alpha$ governing the deceleration of the clouds, and a parameter $\log(f)$ which is the logarithm of the fractional uncertainty $f$ that the formal measurement error on the clouds' past velocities may be underestimated. 
\label{fig:cornerplot} }
\end{figure}

Given these priors and our likelihood function, we adopt the default parameters of the \texttt{dynesty} dynamic nested sampler. A cornerplot showing the well-characterized 1D and 2D marginal distributions of the model parameters is provided in Figure \ref{fig:cornerplot}. We derive the median (50th percentile) of the samples, and define the upper and lower error bounds as the difference between the 50th and 84th percentiles of the samples, and the 16th and 50th percentiles of the samples, respectively. All the uncertainties we report are \textit{statistical} uncertainties, which are lower limits on the actual uncertainties. There is an additional systematic uncertainty stemming from the simplicity or our model assumptions, which is unknown but very likely dominates. We discuss the additional sources of systematic uncertainty in \S \ref{subsec:uncertainties}.

We obtain a value of $n_0 \rm = 0.041^{+0.004}_{-0.003} \; cm^{-3}$ for the ambient volume density of hydrogen nuclei, $t_{exp} \rm = -1.22^{+0.07}_{-0.08} \; Myr$ for the explosion time, $\rm \alpha = 14.53^{+4.92}_{-4.66} \; pc \; Myr^{-2}$ for the deceleration term, and $\delta v_{sh} \rm = -13.89^{+2.60}_{-2.35} \; km \; s^{-1}$ for the velocity offset term due to our simplified model. For the logarithm of the fractional uncertainty estimate on the present day cloud velocities we obtain ${\rm log}(f) = 1.81^{+0.26}_{-0.27}$, consistent with the cloud uncertainties in the past being underestimated by $\approx 6-7\times$. Since $\sigma v_{mag} \rm \approx 1 \; km \; s^{-1}$ in Table \ref{tab:clic6d}, the model favors an additional uncertainty on the cloud velocity magnitude measurements (at the time of intersection with the shell in the past) of $\rm \approx$ 6--7 \; km \; s$^{-1}$, likely accounting for additional errors stemming from the simple model governing the cloud's deceleration over time. 

Given the larger value of $\log(f)$, we calculate the reduced chi-squared, $\chi_{\rm red}^2$, of our fit to confirm that our errors are not overestimated:

\begin{equation} \label{eq:redchi2}
\chi^2_{\text{red}} = \frac{\chi^2}{\nu} = \frac{1}{N - m} \sum_{i=1}^{N} \frac{(v_{cl_{intersect,i}} - v_{sh_{intersect,i}})^2}{\sigma_{v_{cl,i}}(f)^{2}}
\end{equation}

\noindent where $v_{cl_{intersect,i}}$ and $v_{sh_{intersect,i}}$ are the cloud and shell velocity, respectively, at the time of intersection for the $ith$ cloud (see Table \ref{tab:clic_ages}). The $\sigma_{v_{cl,i}}(f)$ term accounts for the inflated uncertainties, equal to $\sigma v_{mag_{LSR},i} + (f\times \sigma v_{mag_{LSR},i})$. The ${N - m}$ term is the number of degrees of freedom, calculated as the difference between the number of clouds ($N$) and the number of model parameters ($m$). Plugging everything into Eq. \ref{eq:redchi2}, we obtain $\chi^2_{\text{red}} = 1.2$, indicating that model fits the data well and the uncertainties on the data are appropriately estimated. 

We can compare our inferred ambient volume density of hydrogen nuclei $n_0$ to complementary constraints on the present day electron density inside the Local Bubble. \citet{linsky2021} found a typical electron density of $n_e = 0.012 \; \rm cm^{-3}$, which they obtained by averaging dispersion measure data towards five pulsars along lines of sight that intersect the Local Bubble. In our model, we infer the total volume density of hydrogen nuclei (rather than the electron density). Since we do not know the ionization fraction (either within the present day Local Bubble or in the ambient interstellar medium prior to the supernova explosion), let us make the simplifying assumption that the bubble is fully ionized for the sake of comparison, such that $n_e = n_{H^+}$.  In this scenario, the total volume density of hydrogen nuclei would also be equal to the electron density: $n_0 \approx n_e \approx n_{H^+}$. Therefore, our model predicts that the density of the bubble in the present day is roughly $3-4\times$ lower than the ambient density prior to the most recent supernova exploding $\approx 1.2 \; \rm Myr$ ago.

We can also compare our estimate for the time of the supernova explosion $t_{exp}$ to the influx of iron deposits in Earth's crust. $^{60}\rm Fe$ is a radioactive isotope (half-life of 2.6 Myr), which is produced predominantly in massive stars and found in supernova ejecta \citep{Wallner_2015}. As the Sun has been located interior to the supernova-driven Local Bubble for the past several million years, the Earth has therefore been exposed to $^{60}\rm Fe$, either by being exposed to waves of supernova ejecta directly, or by traversing clouds of $^{60}\rm Fe$-enriched dust. \citet{wallner2021} found two peaks in the $^{60}\rm Fe$ record in samples of the Pacific Ocean crust: one between $\approx 1.7-3.2\; \rm Myr$ ago and one at $\approx 6\; \rm Myr$. The Sun's passage through a potentially $^{60}\rm Fe$-enriched CLIC could be responsible for the much smaller enhancement in the $^{60}\rm Fe$ record over the past tens of thousands of years. However, it does not explain the significant peak $\approx 1.7-3.2\; \rm Myr$ ago. \citet{Opher_2024} proposed that the Sun's crossing of the Local Lynx Cold Cloud could explain for the $\approx 1.7-3.2\; \rm Myr$ $^{60}\rm Fe$ peak. However, one alternative explanation is that the Sun interacted with one or more waves of supernova ejecta from the Sco-Cen association directly. If so, our estimate of $t_{exp} = -1.2 \; \rm Myr$ is not inconsistent with the $\approx 1.7-3.2\; \rm  Myr$ enhancement, given the very broad $^{60}\rm Fe$ peak and typical absolute age uncertainties of $\approx 0.3-0.5$ Myr on the \citet{wallner2021} measurements. 

\begin{figure*}
\includegraphics[width=1.0\textwidth]{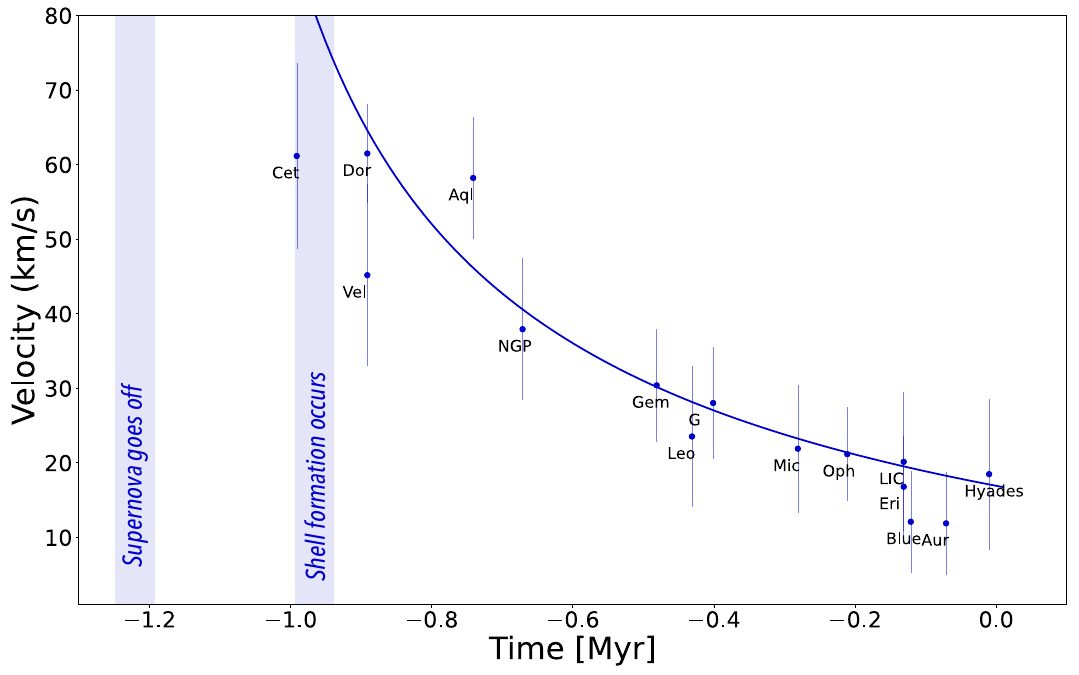}
\caption{Velocity evolution of the supernova shell and the CLIC clouds. The thick blue line shows the velocity of the shell as a function of time ($v_{sh} + \delta v_{sh}$ in the context of Figure \ref{fig:likelihood} and Equation \ref{eq:logl}), defined by the median of the samples from the \texttt{dynesty} run. The blue dots show the velocity of the CLIC clouds when they intersect the surface of the expanding shell, providing a prediction for the ages of the individual clouds. The errorbars on the blue dots show the error on the cloud velocities inferred as part of our modeling.  Cloud formation is predicted to occur roughly on or after the onset of shell formation at $t_{sf} = -0.97 \; \rm Myr$, or $\approx 250,000$ years after the supernova exploded at $t_{exp} = -1.22 \; \rm Myr$. 
\label{fig:shell_evolution} }
\end{figure*}

\section{Discussion} \label{sec:discussion}

\begin{figure*}
\centering
\includegraphics[width=1.0\textwidth]{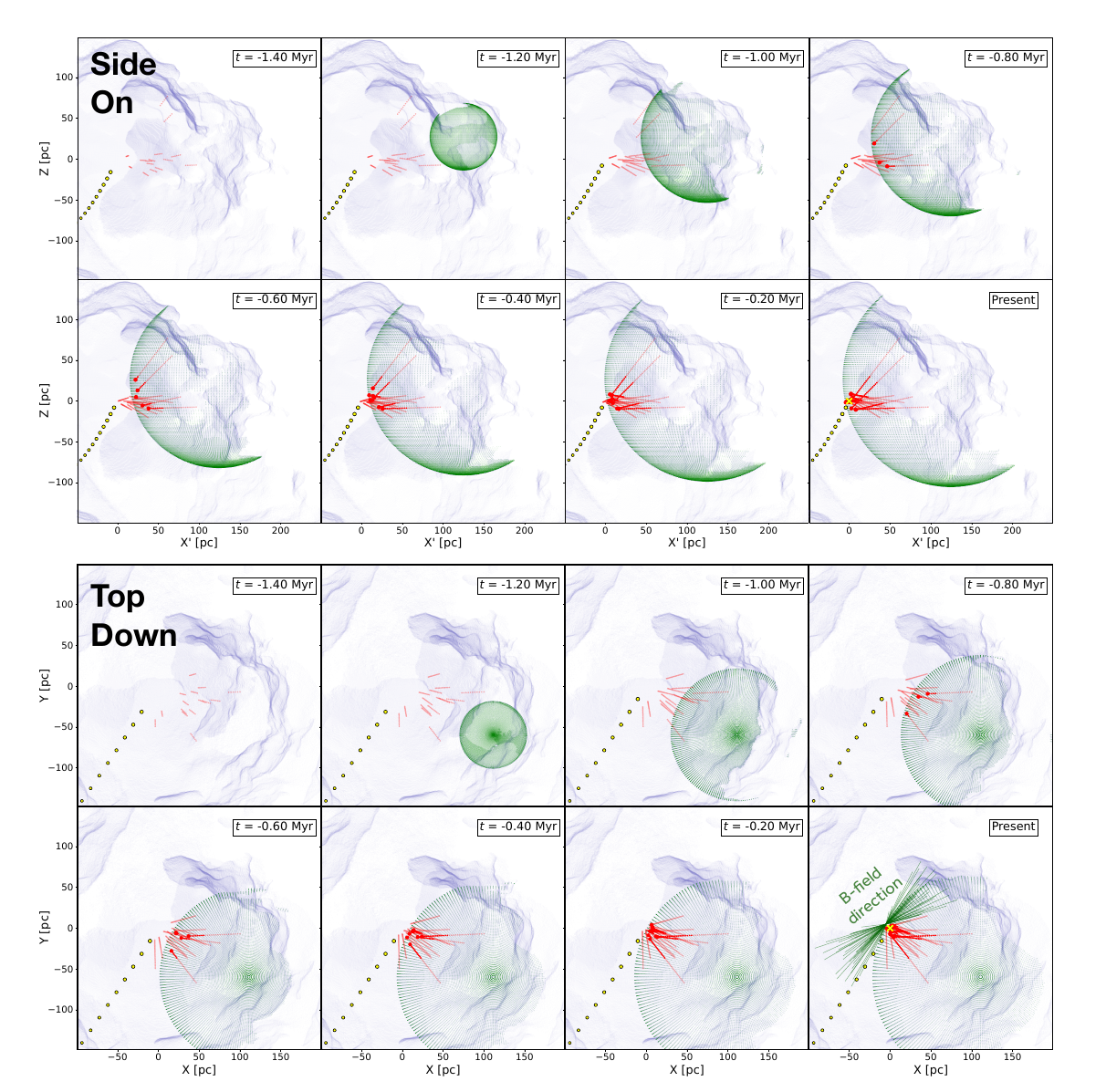}
\caption{Proposed scenario for the formation of the CLIC on the surface of an expanding shell, stemming from the most recent supernova explosion in the UCL cluster. The top set of panels shows a side-on view of the evolution (rotated $20^\circ$ counter-clockwise for an observer facing the Galactic center, denoted by $\rm X'$), while the bottom set of panels shows a top-down view.  Each sub-panel shows the CLIC (red points), the Sun's trajectory (yellow points), and the expanding supernova shell (green sphere) at specific time snapshots, spanning $t\rm = -1.4 \; Myr$ to $t\rm=0 \; Myr$ (present day). A model for the Local Bubble at $t\rm=0 \; Myr$ \citep[purple surface;][]{ONeill_2024} is also overlaid and is presumed to not have changed significantly over the past $\rm \approx 1.4 \; Myr$, since the cavity was largely carved out by previous supernovae beginning $\rm \approx 14 \; Myr$ ago \citep{zucker2022}. The supernova exploded at $t_{exp}=-1.22 \; \rm Myr$. Prior to their birth, the trajectories of the individual CLIC clouds are shown with semi-transparent red traces meant to guide the eye. After their birth in the expanding shell, the positions of the CLIC clouds are shown with large red dots. The present day top-down sub-panel additionally shows constraints on the local interstellar magnetic field direction \citep[thin green lines;][]{Frisch2022}, indicating that the magnetic field is roughly tangent to the shell. An interactive version of this figure is available \href{https://faun.rc.fas.harvard.edu/czucker/Paper\_Figures/Interactive\_CLIC\_Formation.html}{here.}
\label{fig:cloud_formation} }
\end{figure*}

\subsection{Implications of Shell Modeling for the Properties of the CLIC} 

Here we discuss the implications of our modeling for the properties of the CLIC, including the clouds' predicted ages (\S \ref{subsec:ages}), the predicted swept-up column density of the shell (\S \ref{subsec:column}), and the predicted cooling time and temperature of the shell (\S \ref{subsec:temperature}). In \S \ref{subsec:magnetic}, we discuss our model in the context of existing measurements of the magnetic field in the local interstellar medium. Finally, we discuss caveats of our analysis and potential avenues for refinement in the future in \S \ref{subsec:uncertainties}.

\subsubsection{Predicted Age of the CLIC} \label{subsec:ages}

In Figure \ref{fig:shell_evolution} we show the model for the velocity evolution of the shell ($v_{sh} + \delta v_{sh}$ in the context of Figure \ref{fig:likelihood} and Equation \ref{eq:logl}) given the median of the samples from the \texttt{dynesty} run. We track the shell's evolution from the time the supernova explosion occurred ($ t_{exp}=-1.22 \; \rm Myr$) to the present day ($ t=0 \; \rm Myr$). Clearly, we predict that the CLIC formed after the supernova went off 1.2 Myr ago, providing an upper limit on the age of the complex. However, the expanding shell model also provides a more nuanced estimate for the ages of individual clouds, constrained by the point in time in which the shell intersects the dynamical tracebacks of the clouds and presumably led to their formation on its expanding surface. Alongside the shell's velocity evolution, we also overlay the predicted birth times of the clouds, labeled by name. The predicted ages of the clouds, along with the velocity of the clouds and the corresponding shell velocity at their time of birth, are summarized in Table \ref{tab:clic_ages}. After correcting the shell's velocity for the inferred $\delta v_{sh}$ parameter (a term that corrects for an overall kinematic shift between the shell and the ensemble of clouds at the time of intersection due to our idealized model) the typical mean dispersion of cloud velocities around the shell is $\rm 6.4 \; km \; s^{-1}$. This mean dispersion is consistent with the our inferred value of $\log(f) = 1.81$ --- the fractional underestimation of the error on the cloud velocity magnitude --- which favors an additional uncertainty on the cloud velocity magnitudes of $\approx 6-7 \; \rm km \; s^{-1}$. Computing the median dispersion, we find a typical value of $\rm 3.1 \; km \; s^{-1}$, suggesting that the mean dispersion is sensitive to outliers like Vel, Cet, and Dor which show greater discrepancy with the shell velocity around the time of shell formation. Vel, Cet, Dor are outliers as they are based on only 7, 5, and 4 lines of sight, so we expect their velocity vectors to be more uncertain. 

As shown in Figure \ref{fig:shell_evolution}, our model implies a total spread in ages of the clouds of almost a million years, with the oldest clouds in the CLIC being Cet, Vel, and Dor (with a typical age of $\rm \approx 900,000\; yr$) and the youngest clouds being the LIC, Eri, Blue, Aur, and the Hyades (with a typical age of $\rm \approx 100,000\; yr$). Critically, all clouds are predicted to form roughly on or after the onset of shell formation. Recall from Equation \ref{eq:shell_formation} that the time of shell formation is dependent on the ambient volume density of hydrogen nuclei $n_0$ \citep{Kim_Ostriker_2015}.

\begin{equation}
t_{sf} = 4.4 \times 10^{4} \; {\rm yr} \; \big(\frac{n_0}{1 \; {\rm cm^{-3}}}\big)^{-0.55}.
\end{equation}

Given our inferred ambient density of $n_0=0.041^{+0.004}_{-0.003} \; \rm cm^{-3}$, shell formation is predicted to occur between  $ t\approx-0.98$ and  $t\approx-0.95 \; \rm Myr$, or $\rm \approx 240,000-270,000 \; \rm yr$ after the supernova exploded at $t_{exp}=-1.22 \; \rm Myr$. The first cloud to be born, Cet, is predicted to have formed at $t \approx -0.99 \; \rm Myr$, consistent with the prediction for shell formation and the onset of the pressure-driven snowplow phase. Snapshots of the shell's evolution, showing the formation of individual CLIC clouds at different epochs given their predicted ages, is shown in Figure \ref{fig:cloud_formation}. Since the predicted time of formation of the oldest cloud (Cet) at $t \approx -0.99 \; \rm Myr$ and $t = 0 \; \rm Myr$ (present day), the shell has expanded by roughly 50 pc, which is a few times the present thickness of the CLIC. 

Analytic models also provide a framework for estimating the lifetime of an expanding shell. One typical assumption is that a supernova remnant ``fades away" when the velocity of the shell approaches the typical sound speed $c_s$ of the interstellar medium. This ``fadeaway" time is given in \citet{draine_2011} as: 

\begin{equation}
t_{fade} \approx \bigg(\frac{2 \; r_{sf}}{7 \; t_{sf} \; c_s}\bigg)^{\frac{7}{5}} t_{sf}
\end{equation}

\noindent where $t_{sf} \rm = 250,000 \; yr$  is the time of shell formation, and $r_{sf} \rm = 84 \; pc$ is the radius at the time of shell formation. Assuming a typical sound speed of $\rm \approx 10 \; km \; s^{-1}$, $t_{fade} \approx 5 \; \rm Myr$. Given the tenuous nature of these diffuse clouds and the hostile EUV-irradiated environment within the Local Bubble, it is reasonable to consider $t_{fade}$ to be an upper limit on the predicted survival time of the CLIC as well as the shell. Indeed, \citet{Provornikova_2011} performed dynamical modeling to estimate the lifetimes of cold neutral clouds surrounded by hot plasma, focusing on clouds with properties similar to the CLIC that are embedded inside conditions typical of the Local Bubble. \citet{Provornikova_2011} found typical lifetimes ranging between $0.2–1.5$ Myr, which is consistent with our predicted ages from Table \ref{tab:clic_ages} and again suggests much shorter lifetimes than $t_{fade}$. 

\subsubsection{Predicted Shell Column Density } \label{subsec:column}

As a consistency check for our model, we can compare the predicted column density of the shell with extant constraints on the column densities of individual clouds in the CLIC. Assuming spherical shell expansion, the column density of the shell that has been swept up is given by:
\begin{equation} \label{eq:surface_density}
N_{sw} = \frac{1}{3} \, n_0 \, r_{sh},
\end{equation}

\noindent where $n_0$ is the ambient density in which the shell is expanding into and $r_{sh}$ is the current shell radius. Adopting our median ambient density of $n_0 \rm = 0.041 \; cm^{-3}$ and our present day shell radius of $r_{sh} = \rm 132 \; \rm pc$, we obtain a column density of $N_{sh} = 5.5\times10^{18} \; \rm cm^{-2}$. Examining the atomic hydrogen column densities towards 33 lines of sight for stars within $\rm \approx 60 \; pc$ from the Sun with stellar Ly$\alpha$ emission lines, \citet{Wood_2005} obtain a typical average atomic hydrogen column density of $\rm \approx 1.4 \times 10^{18} \; cm^{-2}$ for the local interstellar medium, which should be a lower limit on the total hydrogen column density given evidence that the clouds are partially ionized ($X(H^{+}) \approx 0.2$). Thus the observed and predicted column densities agree to within a factor of $\approx 3$. 

While the predicted column density of the shell in the immediate solar vicinity is broadly consistent with observations of the CLIC, mass should also have been swept up elsewhere on the surface of the shell. While no known diffuse clouds exist in the Local Bubble beyond $\rm 15 \; pc$, about 20\% of the velocity components from \citet{redfield2008} cannot be assigned to any of the 15 clouds in the CLIC. Many of these velocity components may represent more distant clouds that cover smaller fractions of the sky and are probed by too few background sources. The absorption associated with some of these components may represent clouds forming elsewhere on the surface of the shell. Our model thus provides a prediction on where other clouds within the Local Bubble may be forming, and underlines the need for a larger sample of background stars with {\it HST} spectroscopy that may show evidence of other parts of the shell in absorption.

\begin{deluxetable}{cccc}
\tablecaption{Predicted Ages of the CLIC \label{tab:clic_ages} }
\colnumbers
\tablehead{\colhead{Cloud} & \colhead{$v_{sh_{intersect}}$} & \colhead{$v_{cl_{intersect}}$} & \colhead{Age}\\
\colhead{} & \colhead{$\rm km \; s^{-1}$} & \colhead{$\rm km \; s^{-1}$} & \colhead{Myr}}
\startdata
LIC & 19.5 & 20.1 & 0.13 \\
G & 27.0 & 28.0 & 0.40 \\
Blue & 19.3 & 12.1 & 0.12 \\
Aql & 46.2 & 58.2 & 0.74 \\
Eri & 19.5 & 16.8 & 0.13 \\
Aur & 18.3 & 11.9 & 0.07 \\
Hyades & 16.9 & 18.5 & 0.01 \\
Mic & 23.2 & 21.9 & 0.28 \\
Oph & 21.4 & 21.2 & 0.21 \\
Gem & 30.2 & 30.4 & 0.48 \\
NGP & 40.6 & 37.9 & 0.67 \\
Leo & 28.2 & 23.5 & 0.43 \\
Dor & 64.6 & 61.5 & 0.89 \\
Vel & 64.6 & 45.2 & 0.89 \\
Cet & 87.7 & 61.2 & 0.99 \\
\enddata
\tablecomments{Implications of our shell modeling for the ages of clouds in the CLIC and their inferred velocities at the time of formation. (1) Name of the cloud (2) Velocity ($v_{sh} + \delta v_{sh}$) that the supernova shell has when it intersects the cloud. (2) Velocity that the cloud has when it intersects the supernova shell. Since we assume the clouds formed at this time of intersection, $v_{cl_{intersect}}$ is predicted to be the initial velocity of the clouds at birth.  (4) The modeled age of the cloud, determined by the time of intersection between cloud and shell. A machine readable version of this table is available online at the Harvard Dataverse (\href{https://doi.org/10.7910/DVN/LK0XFS}{doi:10.7910/DVN/LK0XFS}).}
\end{deluxetable}

\subsubsection{Predicted Cooling Time and Shell Temperature} \label{subsec:temperature}

As an additional consistency check of our model, we can compare the inferred time since shell formation with the cooling time of the gas, $t_{cool}$, needed to cool to the observed temperature of the CLIC. According to \citet{redfield_2004}, the weighted mean temperature of the CLIC is $\rm 6680 \pm 1490 \; K$, with a maximum temperature of $\rm \approx 12,000 \; K$. Since atomic fine structure line cooling and $\rm Ly \; \alpha$ cooling become dominant at $T < 10^{4} \; \rm K$, we only consider here the time for the shell to radiatively cool to the temperatures of the warmest clouds in the CLIC, between $\rm 8,000-12,000 \; K$.

Following \citet{Kim_Ostriker_2015}, the shell is predicted to have a temperature $T_{sf}$ at the time of its formation of
\begin{equation}
T_{sf}  = 5.67 \times 10^5 \; {\rm K} \; n_0^{0.26}.
\end{equation}

\noindent Again adopting our inferred ambient density of $n_0 = 0.041 \; \rm cm^{-3}$, we obtain a temperature of $T_{sf} = 2.5 \times 10^{5} \; \rm K$. To model the radiative cooling time of the shell since its formation $t_{sf} = 0.97$ Myr ago, we adopt the same formalism for the cooling time as \citet{Kim_Ostriker_2015}

\begin{equation} \label{eq:cooling_time}
t_{cool} = \frac{\beta\,k\,T_{sf}}{(\gamma - 1) \, n_{sh} \, \Lambda},
\end{equation}
where $\gamma = \frac{5}{3}$ is the specific heat ratio, $k$ is the Boltzmann constant, $n_{sh}$ is the shell density at the time of shell formation, $\Lambda$ is the cooling function, and $\beta$ is a dimensionless parameter accounting for the definition of the cooling function. We adopt $n_{sh} = 0.11 \; \rm cm^{-3}$ at the time of shell formation, which is computed from the predicted mass of the shell at its formation ($M_{sf} = 3850 \; \rm M_\sun$)\footnote{The mass at shell formation is given in \citet{Kim_Ostriker_2015} as $M_{sf} = 1680 \; {\rm M_\sun} \times n_0^{-0.26} = 3850 \; \rm M_\sun$ assuming our inferred ambient density $n_0 = 0.041 \; \rm cm^{-3}$.} and radius at shell formation ($r_{sf} = 84 \; \rm pc$) assuming a shell thickness of ($\delta r_{sh} = 10 \; \rm pc$). This shell thickness is a reasonable assumption given the $\approx$ 10 pc distance spread of the CLIC clouds in the present day. For $\Lambda$, we adopt the standard cooling function for solar metallicity gas from \citet{gnedin_2012}, which assumed pure collisional ionization equilibrium and solar-metallicity. Following the definition of $\Lambda$ from \citet{gnedin_2012}, who use the total baryon number density instead of the hydrogen number density, $\beta$ can be approximated as 1.173.

\begin{figure}
\includegraphics[width=0.47\textwidth]{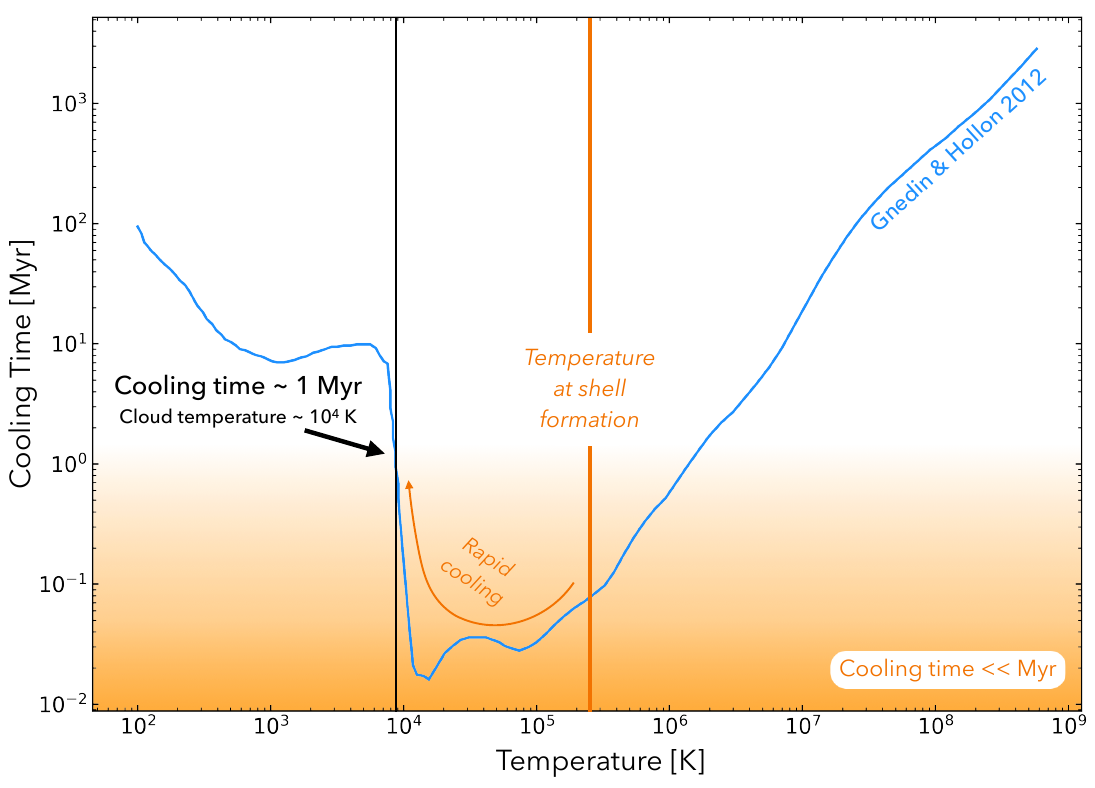}
\caption{Cooling time versus temperature given the cooling function of \citet{gnedin_2012}. After the shell forms at a temperature of $\rm 2.5\times10^{5} \; K$, the shell will rapidly cool to $\rm 10^{4} \; K$ and remain at this temperature on the order of $\rm \approx 1 \; Myr$ or more. Given that shell formation occurred $t_{sf}=0.97 \; \rm Myr$ ago and the CLIC's current temperature of $\rm 5,000-10,000 \; K$, the predicted cooling is in agreement with our model for the shell evolution and the observed thermal properties of the CLIC.\label{fig:cooling_itme}}
\end{figure}

Figure \ref{fig:cooling_itme} shows the cooling time versus temperature computed using Equation \ref{eq:cooling_time} with the \citet{gnedin_2012} cooling function and our adopted value of $ n_{sh} = 0.11 \; \rm cm^{-3}$. After the shell forms at a temperature of $T_{sf} = 2.5 \times 10^{5} \; \rm K$, it will cool rapidly to $\rm \approx 10^{4} \; K$ (the approximate observed temperature of the CLIC) and remain at this temperature plateau for at least a megayear. Thus, the predicted radiative cooling time is in agreement with our inferred time since shell formation $t_{sf}=0.97 \; \rm Myr$ ago, validating our model.  

\subsubsection{Shell Projection and CLIC Distribution}

\begin{figure*}
\centering
\includegraphics[width=1.0\textwidth]{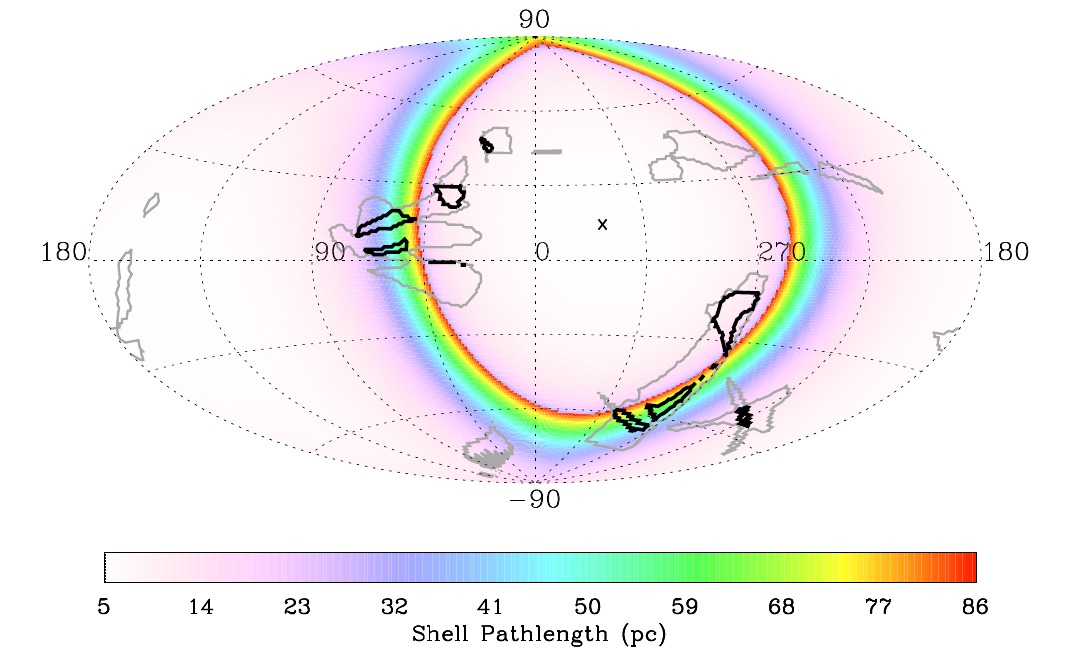}
\caption{Projection of the shell on the sky in Galactic coordinates. Assuming we are currently within the shell and that it has a thickness of 10 pc, the color shading gives the path length through the shell. The center of the shell is indicated by the $\times$ symbol. The minimum path length of 5 pc is along the radial direction from the shell center. The maximum path length ($>$50 pc) is seen along a thin ribbon tangent to the shell surface. The contours indicate directions in which $\ge$3 (gray) and $\ge$4 (black) LISM clouds are traversed along the line of sight based on the \citet{redfield2008} kinematic model of the CLIC. There appears to be a spatial correlation between high path lengths of the shell model and regions with high numbers of LISM clouds, though it is not uniform around the ribbon. Such a correlation supports the idea that the origin of the CLIC is associated with a recent supernova shell. \label{fig:shellproj}}
\end{figure*}

In Figure~\ref{fig:shellproj} we show the projection of the shell onto the plane of the sky. We assume a thickness of the shell of $\delta r_{sh} = 10 \; \rm pc$ based on the CLIC distribution --- see Figure 11 in \citet{frisch2011} and updates in CLIC densities from \citet{Linsky_2022}. We place the Sun in the center of the shell. As a result the path length is 5\,pc along the radial direction from the shell center to the Sun. The longest path lengths through the shell are perpendicular to the radial direction and appear as a thin ribbon on the sky as shown in Figure~\ref{fig:shellproj}. The large radius of the shell results in a relatively large path length through the shell in the tangent direction, exceeding 50\,pc. If warm, CLIC-like clouds form in the shell, we would expect the distribution of clouds to correlate with the path length of the shell. The contours in Figure~\ref{fig:shellproj} show locations where the number of traversed clouds is highest ($\ge$3 in gray and $\ge$4 in black) which indeed do correlate with the largest path lengths through the shell, though it is not uniform around the ribbon. This spatial structure was first discussed in \S5.1 of \citet{redfield2008} and identified as an interstellar ``Ring of Fire", where it appeared to be a region where interactions of local interstellar clouds were high and potentially impacting the cloud morphology. The high-path-length regions may be interesting locations to search for more distant CLIC-like clouds that are forming elsewhere in the expanding shell. Such a survey may also address why the high number of clouds is not uniform around ribbon, but concentrated in select regions.

\subsection{CLIC and the Local Interstellar Magnetic Field} \label{subsec:magnetic}

It is well known from both simulations \citep[e.g.,][]{Ferriere1991,Ntormousi2017} and observations \citep[e.g.,][]{Soler2018, Tahani2023} that magnetic field lines will be swept up alongside ambient interstellar gas in expanding supernova-driven bubbles, creating field lines that are tangent to the shell's surface and perpendicular to the shell's expansion velocity. Accordingly, we would expect the direction of the interstellar magnetic field in the CLIC to be oriented perpendicular to its direction of motion.

This idea has been explored extensively in the literature in the context of the CLIC's proposed location embedded within the rim of the S1 supershell of Loop I \citep[see e.g.,][]{frisch2017, frisch2011, frisch2012, frisch2015, wolleben2007}. \citet{Frisch_2015_Polarization} compiled a sample of a few hundred stars within 40 pc from the Sun that possess both measurable linear polarization and known distances from Hipparcos. \citet{Frisch_2015_Polarization} found that the direction of the interstellar magnetic field is oriented $76.8{^\circ}^{+23.5^\circ}_{-27.6^\circ}$ (nearly perpendicular) to the bulk velocity vector of the CLIC, in agreement with predictions for swept up field lines.  This orientation of the interstellar magnetic field has also been shown to agree with the magnetic field derived from the so-called Interstellar Boundary Explorer (IBEX) ``ribbon" of energetic neutral atoms \citep{mccomas_2009} created by the interaction between the interstellar magnetic field and the heliosphere \citep[see discussion in][]{Frisch2022}. \citet{Frisch2022} further found that the partially ionized CLIC clouds couple to the interstellar magnetic field. They argued that at least three of the clouds in the CLIC (Dor, G, Blue) show cloud boundaries \citep[derived from][]{redfield2008} that are aligned with this field, tracing the potential shock interface that formed from collisions of these three clouds.

The same physical picture for the orientation of the interstellar magnetic field presented in \citet{frisch2015} and related work remains highly relevant to this work, with the exception that the shell discussed here is related not to Loop I, but rather to the Local Bubble. New 3D spatial constraints on the architecture of nearby OB associations and superbubbles from Gaia \citep{zucker2022, pelgrims2020, ONeill_2024, ONeill_2024b} indicated that the Sco-Cen association --- argued to have formed Loop I in \citet{frisch2015} --- lies just interior to the wall of the Local Bubble (see Figure \ref{fig:UCL_CLIC_LB}). Moreover, new evidence from \citet{panopoulou2021} suggests that Loop I is located at distances $\gtrapprox 100$ pc, consistent with the idea that while Loop I and the Local Bubble may be interacting, they are doing so far beyond the location of the CLIC.

In the present day top-down view shown in Figure \ref{fig:cloud_formation}, we present updated constraints for the orientation of the interstellar magnetic field from \citet{Frisch2022}, which is aligned with Ecliptic coordinates $(\lambda,\beta) \approx 219^\circ \pm 15^\circ, 43^\circ \pm 9^\circ$ for their most significant detection, corresponding to Galactic coordinates of ($l, b) \approx (40^\circ \pm 17^\circ, 56^\circ \pm 17^\circ$). Assuming Gaussian errors, we sample for the local magnetic field orientation in a Galactic frame fifty times and overlay the direction of each sample. Computing the angle between the mean 3D motion of the CLIC and the observed magnetic field orientation from \citet{Frisch2022} for the set of samples, we obtain an angle of $106^\circ \pm 8^\circ$, where an angle of $90^\circ$ is expected if the magnetic field was fully tangent to the shell's surface.  Thus, the observed orientation of the interstellar magnetic field in the CLIC --- roughly perpendicular to our revised model for the swept-up shell --- remains supporting evidence for the supernova-driven origin of the CLIC in an expanding bubble. 

\subsection{Uncertainties on Shell Evolution} \label{subsec:uncertainties}

While our shell evolution model provides a good fit to the spatial and dynamical properties of the CLIC --- with median velocity dispersions around the shell of $\approx \rm 3 \; km \; s^{-1}$ over the ensemble of clouds (see Figure \ref{fig:shell_evolution}) --- it is important to consider the uncertainties and assumptions affecting our analysis, as well as potential avenues for improvement in the future. 

In terms of the observational data, there are uncertainties on the underlying properties of the CLIC that, if better constrained, may lead to more refined estimates of the shell's evolution. For example, all of the clouds' distances --- used to derive the 3D positions of the clouds --- are upper limits to the closest edge of the cloud. Since these distances set the initial trajectories used to determine the crossing time between cloud and shell, improved distances estimates could lead to changes in the modeling. In particular, better distance constraints may decrease the magnitude of the $\delta v_{sh}$ parameter. Currently estimated to be $\approx \rm -14 \; km\;s^{-1}$, $\delta v_{sh}$ accounts for a shift in velocity between the shell and the ensemble of clouds at the point of intersection, which may be due to physical causes (e.g., the shell has a velocity gradient), limitations in the underlying observational data (e.g., the cloud distances), or simplifications in our modeling (e.g., assuming a uniform ambient density when the supernova exploded). 

In addition to the distance uncertainties, a few of the clouds' kinematics are constrained by only a handful of observed lines of sight. A lack of dynamical constraints on some clouds may also explain why the model prefers that the clouds' velocity uncertainties in the past to be underestimated by a factor of six (or $\rm \approx \; 6 \; km \; s^{-1}$). The larger velocity uncertainties compensate for relative outliers like Cet, whose dynamics are defined by only five lines of sight.

Beyond kinematics, there are uncertainties on both the density of clouds and their geometries and surface areas. When modeling the deceleration in \S \ref{subsec:deceleration} we assume spherical geometries for all clouds and a typical density of $\rm 0.1 \; cm^{-3}$. However, \citet{redfield2008} find evidence for a broad range of cloud geometries --- from compact to filamentary --- and argue that cloud densities could be as high as $\rm 0.2 \; cm^{-3}$, although recent work by \citet{Linsky_2022} still supports cloud densities closer to 0.1 cm$^{-3}$. If some clouds are denser and/or more extended, these clouds would experience more deceleration. Failing to account for these effects would lead to greater discrepancies between the cloud velocities and the shell velocity at the time of intersection in Figure \ref{fig:shell_evolution}. Not accounting for the individual variations in cloud deceleration due to their differing morphologies and densities could also explain the very late formation time of clouds like the Hyades, which our model predicts formed only ten thousand years ago --- an estimate that should be considered a lower limit on its actual age. \citet{redfield2009} found that the Hyades, located at the leading edge of the CLIC, shows evidence for increased deceleration in comparison to the rest of the complex. More nuanced modeling of the variation in deceleration from cloud to cloud --- particularly towards the Hyades -- and a full three-dimensional model of the CLIC would thus lead to more accurate determination of the shell's evolution. 

Independent of observational uncertainties on the properties of the CLIC, we also make assumptions in the modeling that deserve further consideration. First, our model assumes that the most recent supernova explosion in UCL occurred somewhere along the \textit{average} trajectory of the cluster. However, the UCL cluster extends roughly 100 pc in size  \citep[e.g.,][]{Ratzenbock2023}, so there is considerable uncertainty in the precise epicenter of the explosion. The same assumption about the average trajectory of UCL over the past few Myr is also made in \citet{neuhauser2020} who determined an explosion time of $t\rm =-1.78 \; Myr$ by constraining when the average UCL position intersected tracebacks of nearby runaway stars and pulsars. If the epicenter of the explosion actually occurred further from the average position of the UCL cluster, this discrepancy could account for the more recent explosion time ($t=-1.22 \; \rm Myr$) obtained in this work, as our model may attempt to trade off distance with time.  

Finally, our model makes a specific assumption about the formation of the CLIC as a function of time -- namely that each cloud inherits the velocity of the expanding shell at the time of its birth, before decoupling and evolving independently from both the shell and from the other CLIC clouds forming within it. While this is a necessary simplification given the limitations of the observational data (on cloud 3D positions, kinematics, densities, shapes etc.) future comparisons with numerical simulations \citep[e.g.,][]{Miniere_2018} tracing the evolution of clouds forming in the expanding dense shell of supernova remnants in their pressure-driven snowplow phase should allow us to revisit our model with more sophisticated assumptions on the precise conditions of their birth. 

\section{Conclusions} \label{sec:conclusions}
In this work, we explore the origin of the CLIC, a set of fifteen kinematically distinct, partially ionized ($\rm \chi(H^{+}) = 0.2$), diffuse ($n \rm \approx 0.1 \; cm^{-3}$), warm ($T = 5,000-10,000 \; \rm K$) clouds that represent the interstellar environment just beyond our heliosphere. To do so, we leverage 3D spatial and 3D dynamical constraints on the CLIC, nearby Str\"{o}mgren spheres, nearby OB associations, and nearby superbubbles to constrain the history of interstellar gas in the immediate solar vicinity. Our conclusions are as follows:

\begin{itemize}

\item We test the hypothesis from \citet{linsky2021} that the origin of the CLIC could be tied to nearby Str\"{o}mgren spheres. We compile a census of Str\"{o}mgren spheres in the solar vicinity and trace the two largest Str\"{o}mgren spheres (around $\rm \epsilon \; CMa$ and $\rm \beta \; CMa$) back in time alongside the CLIC. We find that $\epsilon$ CMa and $\beta$ CMa share a similar backward trajectory, with their ionizing radiation producing a larger combined Str\"{o}mgren sphere with a radius of $R_s=173$ pc. Under the strong assumption of a constant electron density, we find that the CLIC's trajectory has been located interior to this combined Str\"{o}mgren sphere surface over the past several million years. Its interior position, combined with the CLIC's motion transverse to the sphere's trajectory, disfavors (but does not exclude) a purely Str\"{o}mgren sphere origin for the complex. Additionally, we find that Local Bubble may have been born with and has at least partially overlapped with the combined $\rm \epsilon \; CMa$ and $\rm \beta \; CMa$ Str\"{o}mgren spheres since its formation $\approx$14 Myr ago. 

\item We examine evidence for a supernova origin of the CLIC \citep[e.g.,][]{frisch2011, frisch2012}, stemming from the Sco-Cen OB association. We propose that the most recent supernova occurring $\approx$1--2 Myr in the UCL subcluster of Sco-Cen \citep{neuhauser2020} propagated into the pre-evacuated cavity of the Local Bubble, sweeping up tenuous ambient material and giving rise to the CLIC. Given this hypothesis, we fit for the evolution of this supernova remnant using the backward trajectories of the CLIC clouds. 

\item We find that a supernova which went off in the UCL cluster $\approx$ 1.2 Myr ago and propagated into an ambient medium of $n \rm \approx0.04 \; cm^{-3}$ provides the best fit to the backward trajectories of the CLIC clouds. Our model indicates a median dispersion of only $\approx$ 3 km\,s$^{-1}$ between the shell velocity and the ensemble of cloud velocities at the time of cloud formation, validating the underlying assumptions of our model. 

\item All clouds are predicted to form after the onset of shell formation $\approx$ 1 Myr ago and have continued to form over the past $\approx$ 100,000 years, arguing for a young age for these clouds and an age spread of almost a megayear across the CLIC.  

\item We compute the predicted column density and temperature of the shell in the present day and find them to be consistent with the observed properties of the CLIC. The CLIC's magnetic field, oriented perpendicular to its bulk velocity vector, is also consistent with the expected configuration of a magnetic field that has been compressed in an expanding shell, as established by extant studies \citep{frisch2015}. 

\item There is a correlation in the projection of the expanding shell with previously identified regions where most local interstellar clouds are distributed. This correlation supports the argument that CLIC-like clouds are formed by an expanding supernova shell. Distant CLIC-like clouds are most likely to be detected along the thin ribbon tangent to the expanding shell. 

\end{itemize}

Ultimately, our model provides a quantitative framework for the origin of the CLIC, one which offers a consistent explanation for the observed 3D positions, motions, column densities, temperatures, and magnetic field properties of the complex. Future augmentation of the existing catalog of nearby stars with {\it HST} spectroscopy, a three-dimensional model for the CLIC, and improved numerical modeling of mass condensation in expanding supernova shells, should allow further refinement of our model in the coming years.


\begin{acknowledgments}
S.S. and S.R. gratefully acknowledge the National
Science Foundation's support of the Keck Northeast Astronomy
Consortium's REU program through grant AST-1950797. S.R. acknowledges support from the NASA Outer Heliosphere Guest Investigators Program to Wesleyan University for grant No. 80NSSC20K0785.  CZ acknowledges that support for this work was provided by NASA through the NASA Hubble Fellowship grant \#HST-HF2-51498.001 awarded by the Space Telescope Science Institute (STScI), which is operated by the Association of Universities for Research in Astronomy, Inc., for NASA, under contract NAS5-26555. The authors would like to thank Andreas Burkert, Michael Foley, Alyssa Goodman, Jo$\rm \tilde{a}$o Alves, Jonathan Slavin, Martin Piecka, Joshua Peek, Joshua Speagle, and Andrew Saydjari for helpful discussions that contributed to this work. The authors also thank the anonymous referee for constructive feedback that improved the quality of this manuscript. 
\end{acknowledgments}

%

\vspace{5mm}


\software{\texttt{astropy} \citep{2013A&A...558A..33A,2018AJ....156..123A, Astropy2022}, \texttt{glue} \citep{beaumont2015,robitaille2017}, \texttt{galpy} \citep{galpy}}




\bibliography{main}{}
\bibliographystyle{aasjournal}

\begin{appendix}

\section{Calculation of Sizes of Nearby Str\"{o}mgren Sphere} \label{sec:stromgren_appendix}



For each star in our sample (described in \S \ref{sec:stromgren}), we compute the radius of its Str\"{o}mgren sphere, $R_s$, following \citet{ryden_peterson_2020} \citep[based on the classical result from][]{stromgren1939}: 
\begin{equation} \label{eq:stromgren}
R_s = \bigg(\frac{3Q_*}{4\pi \, \alpha(T_e) \, n_e^2}\bigg)^{\frac{1}{3}}, 
\end{equation}

\noindent where $Q_*$ is the number of ionizing photons per second, $\alpha(T_e)$ is the temperature-dependent recombination coefficient, and $n_e$ is the number density of electrons inside the Str\"{o}mgren sphere (assumed to be equal to the number of protons $n_p$). For $\alpha(T_e)$ and $n_e$, we adopt the same values as \citet{linsky2021}, corresponding to $\alpha(T_e) = 4 \times 10^{-13} \; \rm cm^{3} \; s^{-1}$ \citep{Harwit_1988} and $n_e = 0.012 \; \rm cm^{-3}$. The electron density is inferred using the mean electron density along lines of sight to the nearest five pulsars at distances of $156–372$ pc \citep[see Table 1 in][]{linsky2021}.

The number of ionizing photons per second, $Q_*$, is parameterized as:
\begin{equation}
Q_* = \int_{\nu_0}^{\infty} \frac{B_\nu \, 4 \pi r_\star^{2}}{h \nu} \,d\nu,
\end{equation}

\noindent where $B_\nu$ is the Planck function given as:

\begin{equation} \label{eq:planck}
B_\nu =\frac{2h \nu^3}{c^{2} (e^{\frac{h \nu}{kT}} - 1)}. 
\end{equation}

\noindent Here, $r_\star$ is the stellar radius and $\nu_0$ is the frequency corresponding to $\rm \lambda_0 = 912 \; \AA$, which was chosen because only radiation flux at wavelengths shortward of $\rm \lambda = 912 \;  \AA$ is known to photoionize hydrogen. For the white dwarfs, we adopt the average stellar radius of $r_\star = 0.014 \; \rm R_\odot$ reported in \citet{tat1999}. For the bright EUV stars we adopt the standard radii for their spectral type, with the exception of $\rm \epsilon \; CMa$ and $\rm \beta \; CMa$, whose specific radii of $\rm 12.7 \; R_\odot$ and $\rm 8.9 \; R_\odot$ are provided in \citet{underhill1979}. For the white dwarfs, we adopt the temperatures reported in \citet{tat1999} where available, and from the Montreal White Dwarf Database otherwise \citep{Dufour_2017}. For the bright EUV stars, we adopt the standard temperature of their spectral type, with the exception of $\epsilon \; \rm CMa$ and $\beta \; \rm CMa$, whose EUV spectrum are best characterized by blackbody temperatures $T = 17,300 \; \rm K$ \citep{vallerga1995} and $T = 18,700 \; \rm K$ \citep{cassinelli1996}, respectively. 

However, the EUV flux from $\epsilon$ CMa has been examined much more extensively in \citet{vallerga1995}. \citet{vallerga1995} determine an integrated flux for $\epsilon$ CMa of $13150 \; \rm cm^{-2} \; s^{-1}$ between $504-912 \, \rm \AA$ independent of absorption by the cluster of local interstellar clouds, whose hydrogen column density strongly impacts the observed EUV flux. Assuming a modern distance of $d=124 \; \rm pc$ to $\epsilon$ CMa \citep[see Table \ref{tab:stromgren_full} and][]{Linsky_2022}, we obtain a value of $Q=2.43\mathrm{e}+46 \; \rm s^{-1}$, equating to an $R_s = 151 \; \rm pc$. Using the blackbody formalism described above (Eq. \ref{eq:planck}), we obtain a lower value of $R_s = 118 \; \rm pc$ for $\epsilon$ CMa. We include the value based on the observed EUV flux ($R_s=151 \; \rm pc$) in Table \ref{tab:stromgren_full} but note that adopting the smaller value of $R_s = 118 \; \rm pc$ will not alter any conclusions of this work. 

In Table \ref{tab:stromgren_full} we show the Str\"{o}mgren radii for all stars in the sample with an $R_s > 5\; \rm pc$. For each star in Table \ref{tab:stromgren_full}, we additionally include its heliocentric Galactic Cartesian Coordinates ($x,y,z$), and the corresponding 3D space motions ($u,v,w$) with respect to the LSR. For stars with detections in both Gaia and Hipparcos, we prioritize the Gaia astrometry for computing the 3D positions and 3D velocities. We query SIMBAD for each star's radial velocity, as the vast majority of stars do not have a Gaia radial velocity measurement. The underlying astrometric and spectroscopic data are likewise summarized in Table \ref{tab:stromgren_full}. A small number of stars in the sample (WD0501-289, WD1057+719, WD1123+189, WD1159-035, WD1501+664, WD1520+525, WD1634-573) had $Rs > 5 \; \rm pc$ but lacked a parallax, proper motion, and/or radial velocity constraint. Since we could not calculate the 6D phase information for these seven stars, they are excluded from Table \ref{tab:stromgren_full} but may be of interest for follow up studies. 
\begin{deluxetable}{ccccccccccccccccc}
\tablecaption{Properties of Nearby \text{Str\"{o}mgren} Spheres \label{tab:stromgren_full} }
\tabletypesize{\scriptsize}
\colnumbers
\tablehead{\colhead{Name} & \colhead{$\alpha$} & \colhead{$\delta$} & \colhead{$\pi$} & \colhead{$\mu_{\alpha^{*}}$} & \colhead{$\mu_\delta$} & \colhead{RV} & \colhead{$x$} & \colhead{$y$} & \colhead{$z$} & \colhead{$u$} & \colhead{$v$}  & \colhead{$w$} &  \colhead{Astrometry} & \colhead{$\log_{10}(T_{\mathrm{eff}})$} & \colhead{$\log_{10}(Q)$} & \colhead{$R_s$} \\ \colhead{} & \colhead{$^\circ$} & \colhead{$^\circ$} & \colhead{mas} & \colhead{$\rm mas \; yr^{-1}$} & \colhead{$\rm mas \; yr^{-1}$}  & \colhead{$\rm km \; s^{-1}$} & \colhead{pc} & \colhead{pc} & \colhead{pc} & \colhead{$\rm km \; s^{-1}$} & \colhead{$\rm km \; s^{-1}$} & \colhead{$\rm km \; s^{-1}$} & \colhead{} & \colhead{K} & \colhead{$\rm s^{-1}$} & \colhead{pc}}
\startdata
$\epsilon$ CMa & 104.66 & -28.97 & 8.1 & 3.2 & 1.3 & 27 & -61 & -105 & -24 & -3 & -8 & 4 & Hipparcos & 4.24 & 46.39 & 151$^{\rm a}$ \\
$\beta$ CMa & 95.67 & -17.96 & 6.6 & -3.2 & -0.8 & 33 & -101 & -105 & -37 & -12 & -7 & -2 & Hipparcos & 4.27 & 46.09 & 120 \\
WD2211-495 & 333.55 & -49.32 & 17.0 & 11.4 & -66.7 & 26 & 34 & -8 & -46 & 23 & -7 & -10 & Gaia & 4.85 & 44.05 & 25 \\
WD1056+516 & 164.82 & 51.41 & 3.4 & 9.7 & -7.3 & 93 & -142 & 62 & 247 & -22 & 29 & 95 & Gaia & 4.83 & 43.97 & 24 \\
WD0232+035 & 38.78 & 3.73 & 12.9 & 82.6 & 7.4 & 55 & -48 & 12 & -59 & -45 & 5 & -21 & Gaia & 4.83 & 43.94 & 23 \\
WD0621-376 & 95.80 & -37.69 & 13.0 & 66.9 & -2.6 & 32 & -29 & -64 & -28 & 1 & -23 & 16 & Gaia & 4.82 & 43.91 & 23 \\
WD0501+527 & 76.38 & 52.83 & 19.1 & 12.7 & -93.4 & 69 & -47 & 21 & 6 & -62 & 25 & 5 & Gaia & 4.78 & 43.76 & 20 \\
WD0027-636 & 7.49 & -63.42 & 5.1 & -15.7 & -7.2 & 30 & 70 & -93 & -158 & 35 & 3 & -11 & Gaia & 4.77 & 43.72 & 20 \\
WD0455-282 & 74.31 & -28.13 & 8.0 & 51.4 & 12.6 & 70 & -66 & -76 & -74 & -39 & -42 & -8 & Gaia & 4.77 & 43.71 & 19 \\
WD1013-050 & 154.12 & -5.34 & 8.6 & -99.9 & -12.4 & -100 & -32 & -81 & 75 & -2 & 71 & -91 & Gaia & 4.76 & 43.65 & 18 \\
WD0004+330 & 1.88 & 33.29 & 9.9 & -77.9 & -61.7 & 76 & -33 & 81 & -48 & 27 & 83 & -47 & Gaia & 4.75 & 43.60 & 18 \\
WD1314+293 & 199.09 & 29.10 & 16.6 & -158.0 & -107.2 & 54 & 3 & 4 & 60 & -5 & -31 & 67 & Gaia & 4.74 & 43.58 & 18 \\
WD1029+537 & 158.04 & 53.49 & 8.5 & -58.6 & -23.7 & 34 & -65 & 26 & 94 & -35 & 3 & 23 & Gaia & 4.73 & 43.53 & 17 \\
WD2331-475 & 353.51 & -47.24 & 9.4 & -8.6 & -25.1 & 21 & 41 & -19 & -96 & 26 & 1 & -6 & Gaia & 4.74 & 43.54 & 17 \\
WD2309+105 & 348.09 & 10.78 & 13.3 & 139.2 & -19.5 & -16 & 2 & 52 & -53 & -30 & -17 & -2 & Gaia & 4.73 & 43.52 & 17 \\
WD1254+223 & 194.26 & 22.03 & 14.6 & -38.4 & -203.0 & 8 & 4 & -4 & 68 & 36 & -46 & 10 & Gaia & 4.66 & 43.21 & 13 \\
WD2152-548 & 329.09 & -54.64 & 8.0 & 49.8 & -10.4 & -28 & 78 & -29 & -93 & -31 & 12 & 12 & Gaia & 4.65 & 43.16 & 13 \\
WD0715-703 & 108.82 & -70.42 & 10.5 & -108.6 & 113.1 & 7 & 17 & -85 & -37 & -52 & 9 & -26 & Gaia & 4.65 & 43.15 & 13 \\
WD2004-605 & 302.27 & -60.43 & 21.6 & 104.6 & -69.2 & 17 & 35 & -15 & -25 & 5 & 0 & -20 & Gaia & 4.64 & 43.07 & 12 \\
WD0512+326 & 78.85 & 32.68 & 12.3 & -20.7 & 12.6 & -8 & -80 & 9 & -4 & 19 & 22 & 4 & Gaia & 4.63 & 43.04 & 12 \\
HD206697 & 325.68 & 43.59 & 8.9 & 112.4 & 33.3 & -62 & -1 & 112 & -13 & -46 & -49 & -10 & Gaia & 4.60 & 42.88 & 10 \\
WD2111+498 & 318.18 & 50.10 & 19.9 & 93.2 & -16.4 & 94 & -1 & 50 & 0 & -4 & 109 & -9 & Gaia & 4.60 & 42.86 & 10 \\
WD0346-011 & 57.21 & -0.98 & 32.2 & 84.6 & -163.0 & 176 & -23 & -3 & -19 & -116 & -31 & -108 & Gaia & 4.57 & 42.73 & 9 \\
WD0050-332 & 13.32 & -33.00 & 16.9 & -31.1 & 27.5 & 23 & 2 & -5 & -58 & 14 & 24 & -16 & Gaia & 4.56 & 42.62 & 8 \\
WD0347+171 & 57.60 & 17.25 & 21.0 & 127.3 & -22.3 & 37 & -41 & 5 & -22 & -33 & -2 & 5 & Gaia & 4.54 & 42.52 & 8 \\
WD2124+191 & 321.61 & 19.38 & 21.7 & 80.5 & 16.7 & -9 & 14 & 40 & -17 & -7 & 7 & 1 & Gaia & 4.52 & 42.42 & 7 \\
WD2020-425 & 306.00 & -42.41 & 10.1 & -3.4 & 24.9 & -123 & 81 & -2 & -55 & -89 & 29 & 80 & Gaia & 4.47 & 42.10 & 6 \\
HD64511 & 118.77 & 22.00 & 10.7 & -27.4 & -40.3 & 273 & -80 & -28 & 37 & -230 & -81 & 100 & Gaia & 4.48 & 42.13 & 6 \\
WD1234+482 & 189.19 & 47.92 & 7.1 & -86.5 & -61.5 & 28 & -32 & 38 & 131 & -26 & -40 & 45 & Gaia & 4.46 & 41.99 & 5 \\
\enddata
\tablecomments{Summary of the spatial and dynamical properties of significant \text{Str\"{o}mgren} Spheres ($R_s > 5 \; \rm pc$) in the solar vicinity. (1) Name of the star (2) Right Ascension (3) Declination (4) Parallax (5) Proper motion in the right ascension direction (6) Proper motion in the declination direction (7) Radial Velocity (8-10) Heliocentric Galactic Cartesian coordinates (11-13) The 3D space motions along $x$, $y$, and $z$ with respect to the LSR. (14) The source of the astrometry (Gaia or Hipparcos) (15) Log of the effective temperature (16) Log of the total number of ionizing photons emitted by the star per second (17) Str\"{o}mgren radius. A machine readable version of this table is available online at the Harvard Dataverse (\href{ https://doi.org/10.7910/DVN/GZI7CP}{doi:10.7910/DVN/GZI7CP}).}
\tablenotetext{a}{Str\"{o}mgren radius (and corresponding $Q$ value) for $\epsilon$ CMa is based on the EUV flux determination from \citet{vallerga1995}, rather than the formalism in Eq.~\ref{eq:planck}. See \S \ref{sec:stromgren_appendix} for more details.}
\end{deluxetable}

\section{Orbits of Nearby Str\"{o}mgren Spheres} \label{sec:stromgren_orbit_appendix}
\begin{figure*}[ht!]
\includegraphics[width=1.\textwidth]{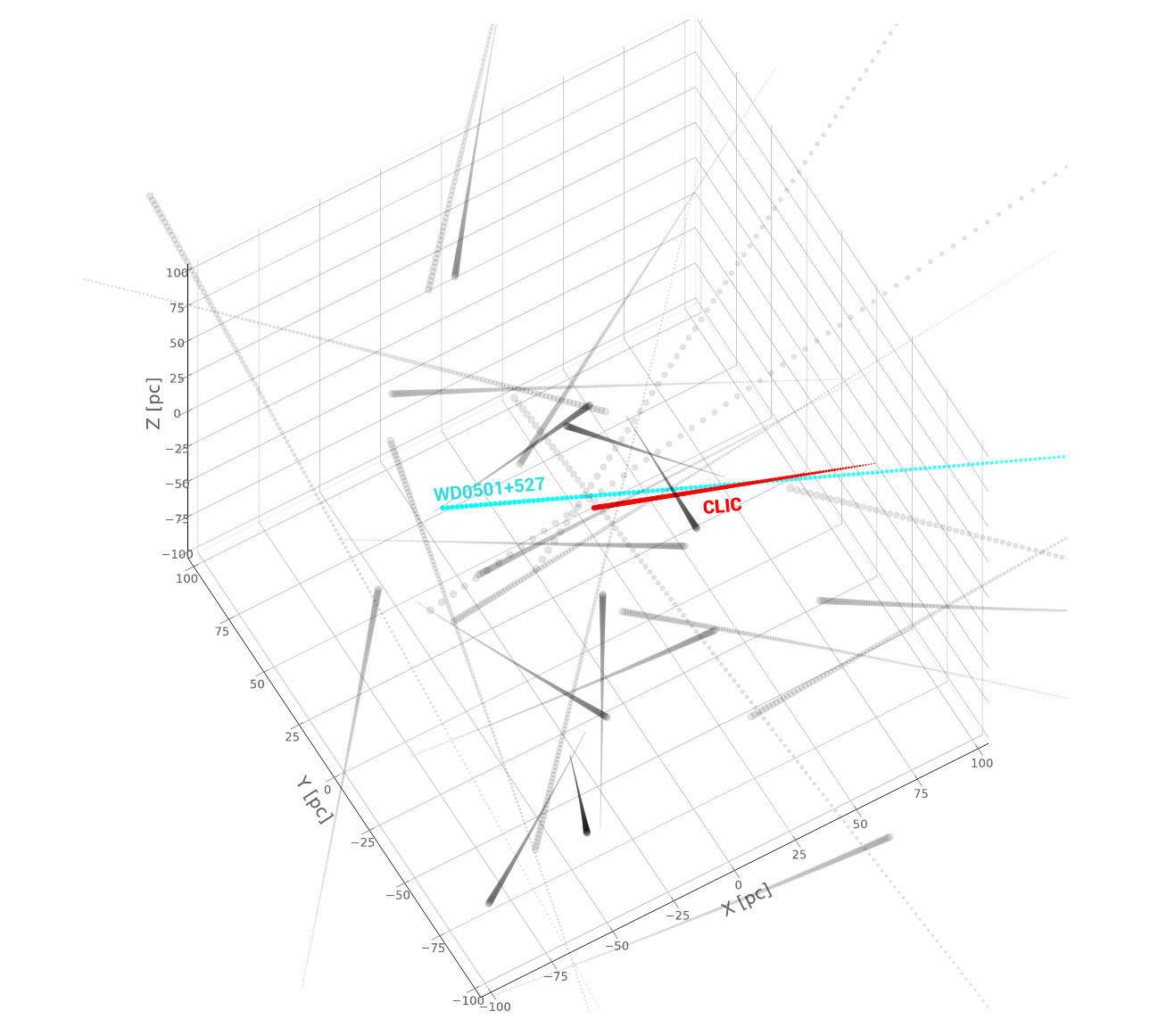}
\caption{The orbital histories (over the past 3 Myr, shown in grey) of all stars hosting Str\"{o}mgren spheres in Table \ref{tab:stromgren_full}. One white dwarf, WD0501+527, shares similar directionality in its orbit (shown in cyan) as that of the CLIC (average shown in red). While the high speed of WD0501+527 ($67 \; \rm km \; \; s^{-1}$) is inconsistent with a CLIC formation scenario tied solely to this hot white dwarf, the surface of the Str\"{o}mgren sphere from WD0501+527 could have interacted with the supernova shockwaves from the UCL cluster and aided the CLIC's formation, as proposed for the largest Str\"{o}mgren sphere in \S \ref{subsec:stromgren_viz}. An interactive version of this figure showing the full animation is available \href{https://faun.rc.fas.harvard.edu/czucker/Paper\_Figures/Stromgren\_Tracebacks.html}{here}.  \label{fig:all_stromgren_orbits}}
\end{figure*}

In Figure \ref{fig:all_stromgren_orbits} we show a 3D view of the orbital histories over the past 3 Myr of all stars with Str\"{o}mgren spheres in Table \ref{tab:stromgren_full}, extending the analysis focused on the largest Str\"{o}mgren sphere from \S \ref{subsec:stromgren_viz}.  The orbital histories are based on the 3D space motions of the stars summarized in Table \ref{tab:stromgren_full} and the procedure for calculating orbits with \texttt{galpy} described in \S \ref{subsec:strom_sample}. As we have shown in \S \ref{subsec:stromgren_viz} (see also \S \ref{sec:stromgren_coevolve}), the largest Str\"{o}mgren sphere in the solar vicinity (stemming from $\epsilon$ CMa and $\beta$ CMa) could not have accounted for the formation of the CLIC alone. However, there are many smaller Str\"{o}mgren spheres that could also play a role in the evolution of the CLIC. For example, if shockwaves from the UCL supernova remnant (\S \ref{sec:SN}) propagated into the surface of a Str\"{o}mgren sphere, this interaction could have led to increased gas compression, cooling, and hydrogen recombination, potentially aiding the CLIC's formation \citep[see e.g.][]{linsky2021}. 

We do find one hot white dwarf --- WD0501+527 with $R_s = 20 \; \rm pc$ --- that shares similar directionality in its ($u,v,w$) space motions in the LSR frame as the mean motion of the CLIC (see the interactive version of Figure \ref{fig:all_stromgren_orbits}, where one can trace each stellar orbit back in time alongside the CLIC). However, the speed of WD0501+527 with the respect to the LSR is $67 \; \rm km \; s^{-1}$, roughly $3\times$ the mean speed of the CLIC and $\approx 1.5\times$ higher than the fastest cloud in the CLIC (the Dor cloud). Therefore, like $\epsilon$ CMa and $\beta$ CMa, WD0501+527 alone could not have contributed to the formation of the CLIC. However, given that the shockwaves from the most recent supernova remnant may have propagated into the surface of the WD0501+527 Str\"{o}mgren sphere over the past 1.2 Myr, it is worth investigating the detailed physics of this possible interaction in future work.

\section{Co-Evolution of the $\epsilon$ CMa and $\beta$ CMa Str\"{o}mgren Spheres} \label{sec:stromgren_coevolve}

\begin{figure*}[ht!]
\includegraphics[width=1.\textwidth]{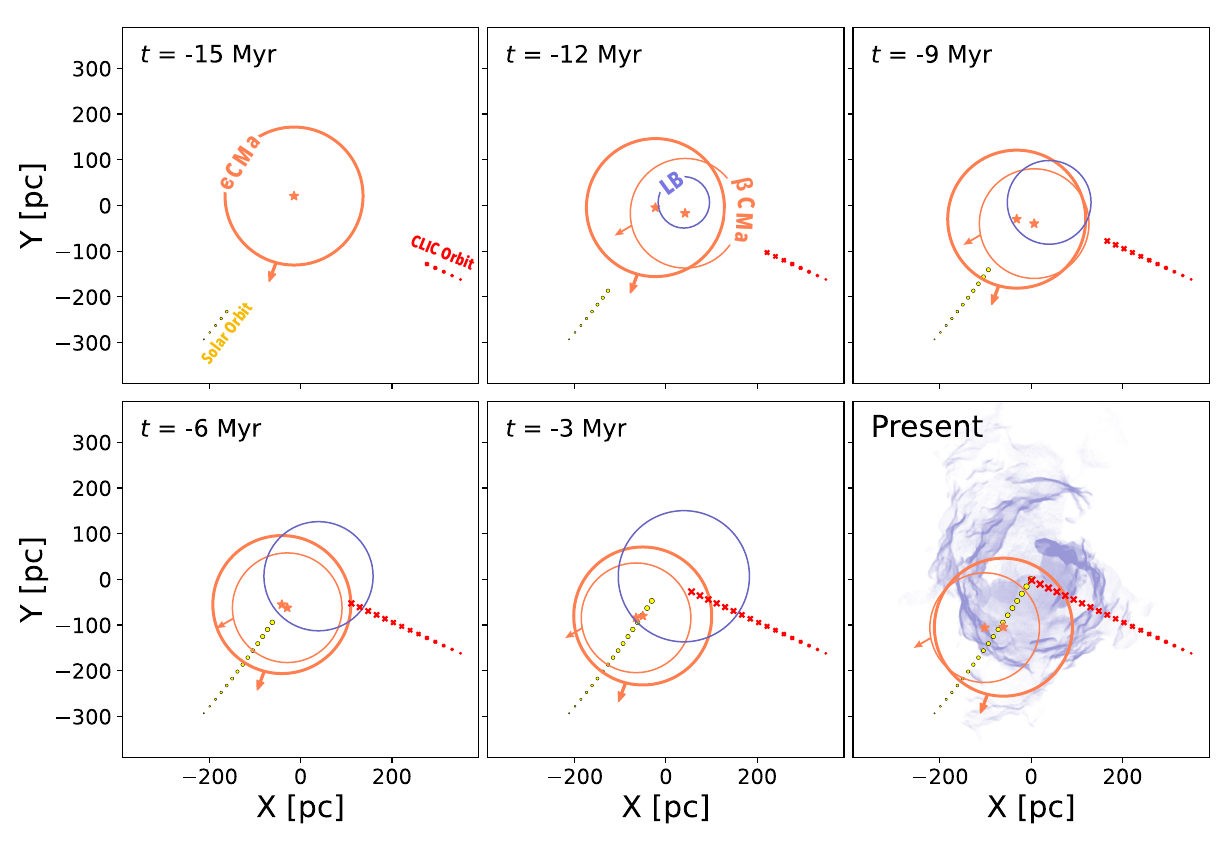}
\caption{Evolution of the individual $\epsilon$ CMa and $\beta$ CMa Str\"{o}mgren spheres (orange rings), the Local Bubble (``LB", purple ring), the CLIC (average trajectory shown in red), and the Sun (trajectory shown in yellow). The evolution of $\epsilon$ CMa and $\beta$ (orange stars) and their respective Str\"{o}mgren spheres are plotted assuming they are completely independent, but given their similar trajectories, their ionizing radiation will form a larger combined sphere (see Figure \ref{fig:stromgren_evolution}). Between $t=-15 \; \rm Myr$ and $t=-12 \; \rm Myr$ both the Local Bubble and $\beta$ CMa formed. \label{fig:stromgren_evolution_independent}} 
\end{figure*}

The Str\"{o}mgren radii, $R_s$, in Table \ref{tab:stromgren_full} (see Appendix \S \ref{sec:stromgren_appendix}) assume that the Str\"{o}mgren spheres evolve independently. However, tracing back the trajectories of $\epsilon$ CMa and $\beta$ CMa over the past 15 Myr, we find that they are roughly co-spatial and share similar $(u,v,w)$ space motions. We demonstrate their correspondence in Figure $\ref{fig:stromgren_evolution_independent}$, which shows the evolution of the relative orientation of $\epsilon$ CMa, $\beta$ CMa, their respective Str\"{o}mgren spheres (assuming no interaction between the two spheres), the Sun's trajectory, the Local Bubble's expanding shell, and the trajectory of the CLIC. In particular, over the past 6 Myr, $\epsilon$ CMa and $\beta$ CMa lie on average $\approx$ 30 pc apart, and pass within 17 pc of each other 3 Myr ago. 

Because $\epsilon$ CMa and $\beta$ CMa share similar trajectories, their Str\"{o}mgren spheres will interact to form a larger Str\"{o}mgren sphere, whose size depends on the combined number of ionizing photons per second produced by the two stars: $Q_{tot} = Q_{\epsilon \; \rm  CMa} + Q_{\beta \; \rm CMa}$. Given $Q_{\epsilon \; \rm  CMa} = 2.43\mathrm{e}+46 \; \rm s^{-1}$ and $Q_{\beta \; \rm  CMa} = 1.24\mathrm{e}+46 \; \rm s^{-1}$, $Q_{tot} = 3.67\mathrm{e}+46 \; \rm s^{-1}$. Plugging $Q_{tot}$ into Eq. \ref{eq:stromgren}, we obtain $R_s = 173 \; \rm pc$ for the combined Str\"{o}mgren sphere stemming from the total ionizing radiation of $\epsilon$ CMa and $\beta$ CMa. We show the evolution of this combined Str\"{o}mgren sphere assuming the average trajectory of $\epsilon$ CMa and $\beta$ CMa in Figure \ref{fig:stromgren_evolution}. 

\end{appendix}

\end{document}